\documentclass[12pt]{article}
\pdfoutput=1

\usepackage{bbm}
\usepackage{amsthm}
\usepackage{mathrsfs}
\usepackage{amsmath, amsfonts, amssymb}
\usepackage{graphicx}
\usepackage{psfrag}
\usepackage[usenames,dvipsnames,svgnames,table]{xcolor}
\usepackage{enumerate}
\usepackage{jheppubcustom}
\usepackage{soul}
\usepackage{subfig}

\newcommand{\be}{\begin{equation}}
\newcommand{\ee}{\end{equation}}

\def\be{\begin{equation}}
\def\ee{\end{equation}}
\def\beq{\begin{equation}}
\def\eeq{\end{equation}}

\newtheorem{teo}{Theorem}[section]    
\newtheorem{lem}[teo]{Lemma}

\def\diam{{\rm diam}}
\def\dist{{\rm dist}}

\newtheorem{theorem}{Theorem}[section]
\newtheorem{lemma}[theorem]{Lemma}

\def\Z{\mathbb Z }

\newcommand{\p}{\partial}

\renewcommand{\[}{\left[}
\renewcommand{\]}{\right]}
\renewcommand{\(}{\left(}
\renewcommand{\)}{\right)}

\def\la{\langle}
\def\ra{\rangle}

\def\simleq{\; \raise0.3ex\hbox{$<$\kern-0.75em
      \raise-1.1ex\hbox{$\sim$}}\; }
   \def\simgeq{\; \raise0.3ex\hbox{$>$\kern-0.75em
      \raise-1.1ex\hbox{$\sim$}}\; }
      
      \newcommand{\figref}[1]{Fig.~\ref{#1}}

\newcommand{\secref}[1]{Sec.~\ref{#1}}

\newcommand{\bs}[1]{\boldsymbol{#1}}

\title{Conformal Correlation Functions in the Brownian Loop Soup}

\author[\bigstar \spadesuit]{Federico Camia,}
\author[\bigstar \clubsuit]{Alberto Gandolfi,}
\author[\bigstar \heartsuit]{and Matthew Kleban}

\emailAdd{federico.camia@nyu.edu, albertogandolfi@nyu.edu, kleban@nyu.edu}

\affiliation[\bigstar]{\it New York University Abu Dhabi, United Arab Emirates}
\affiliation[\spadesuit]{\it VU University, Amsterdam, the Netherlands}
\affiliation[\clubsuit]{\it University of Florence, Italy}
\affiliation[\heartsuit]{Center for Cosmology and Particle Physics,  Department of Physics, New York University, United States of America}

\begin{document}

\abstract{We define and study a set of operators that compute statistical properties of the Brownian Loop Soup, a conformally invariant gas of random Brownian loops (Brownian paths constrained to begin and end at the same point) in two dimensions.   We prove that the correlation functions of these operators have many of the properties of conformal primaries in a conformal field theory, and  compute their conformal dimension.  The dimensions are real and positive, but  have the novel feature that they vary continuously  as a periodic function of a real parameter. We comment on the relation of the Brownian Loop Soup to the free field, and use this relation to establish that the central charge of the Loop Soup is twice its intensity.}

\maketitle

\section{Introduction} \label{intro}

\subsection{The Brownian Loop Soup}

\begin{figure}
\centering
\includegraphics[scale=.5]{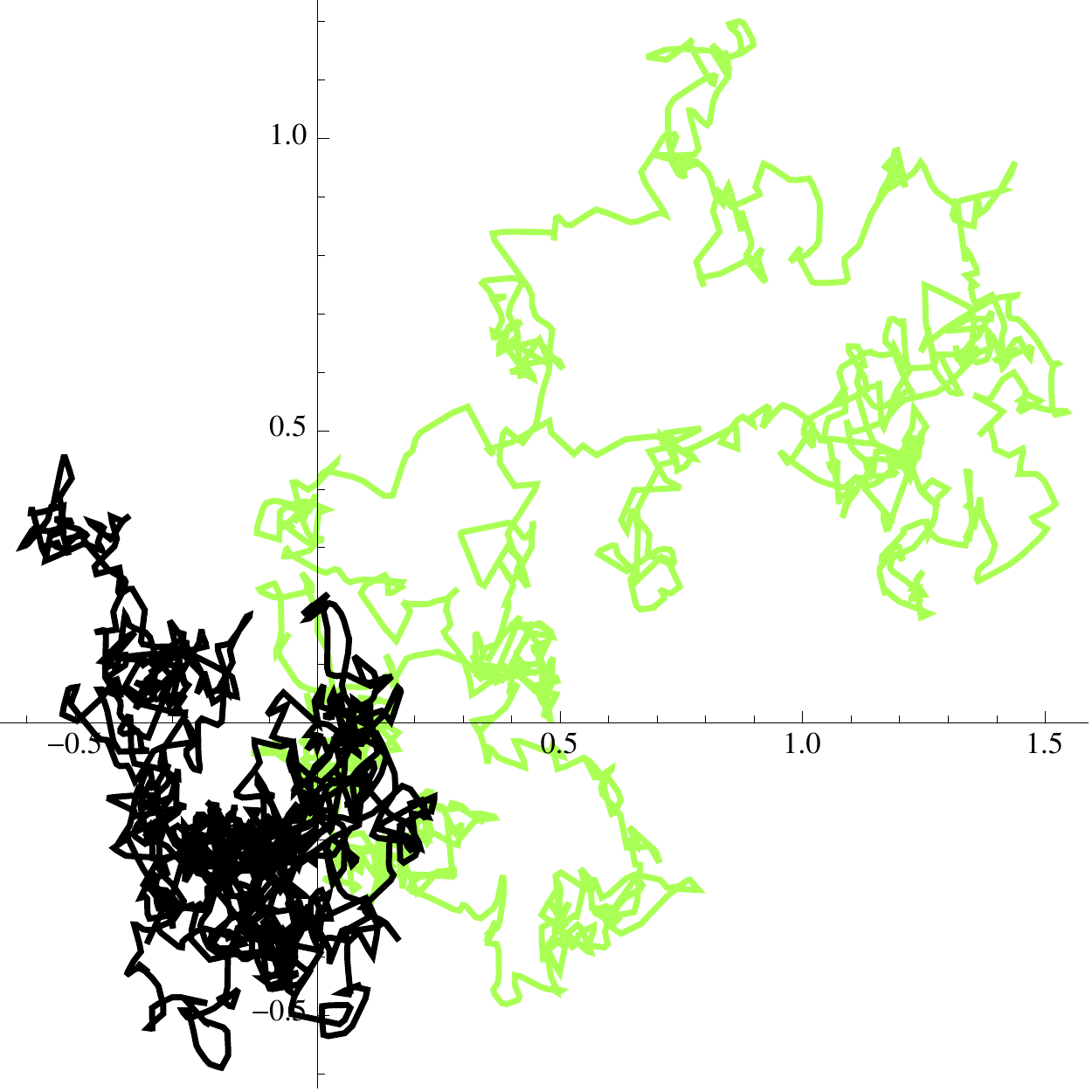}
\caption{Two Brownian loops, each of time length $t=1$.}
\label{loops}
\end{figure}

Take a handful of  loops of various sizes and  sprinkle them onto a flat  surface.  The position where each loop lands is uniformly random, independent of any loops already in place.   Each loop is Brownian -- a Brownian motion constrained to begin and end at the same ``root'' point, but otherwise with no restriction  -- and characterized by a ``time'' length $t$ that is linearly related to its average area (\emph{cf.} \figref{loops}).  The distribution in $t$ is $ \sim dt/t^{2}$, so that there are many more small loops than large, and is chosen to ensure invariance under scale transformations.  The overall density of loops is characterized by a single parameter:  the ``intensity'' $\lambda>0$. This random ensemble of loops is called the Brownian Loop Soup (BLS)
and was introduced in \cite{2003math4419L}.

More precisely, the BLS is a Poissonian random collection of loops in a planar domain $D$ with intensity measure
$\lambda\mu^{loop}_D$, where $\lambda>0$ is a constant and $\mu^{loop}_D$ is the restriction to $D$ of the
\emph{Brownian loop measure}
\begin{equation} \label{brownian-loop-measure}
\mu^{loop} = \int_{\mathbb C} \int_0^{\infty} \frac{1}{2 \pi t^2} \, \mu^{br}_{z,t} \, dt \, d{\bf A}(z) \, ,
\end{equation}
where $\bf A$ denotes area and $\mu^{br}_{z,t}$ is the complex Brownian bridge measure with starting point
$z$ and duration $t$. We note that the Brownian loop measure should be interpreted as a measure on ``unrooted''
loops, that is, loops without a specified ``root'' point. (Formally, unrooted loops are equivalence classes of rooted
loops---the interested reader is referred to \cite{2003math4419L} for the details.)
For ease of notation, the $\mu^{loop}$-measure of a set $\{ \ldots \}$ will be denoted
$\mu^{loop}( \ldots ) \equiv \mu^{loop}(\{ \ldots \}$).

The BLS turns out to be not just scale invariant, but fully conformally invariant. For sufficiently low
intensities $\lambda$, the intersecting loops form clusters whose outer boundaries are
distributed like Conformal Loop Ensembles (CLEs)\cite{MR2979861}. CLEs are the unique ensembles of
planar, non-crossing and non-self-crossing loops satisfying a natural conformal restriction property that
is conjecturally satisfied by the continuum scaling limits of interfaces in two-dimensional models from
statistical physics. The loops of a CLE$_\kappa$ are forms of SLE$_\kappa$ (the Schramm-Loewner Evolution
with parameter $\kappa$ \cite{1999math......4022S}). The CLEs generated by the
BLS correspond to values of $\kappa$ between $8/3$ and $4$. For example, the collection of outermost
interfaces in a planar critical Ising model in a finite domain with plus boundary condition is conjectured to
converge to CLE$_3$ in the scaling limit.

\begin{figure}
\centering
\includegraphics[scale=.6]{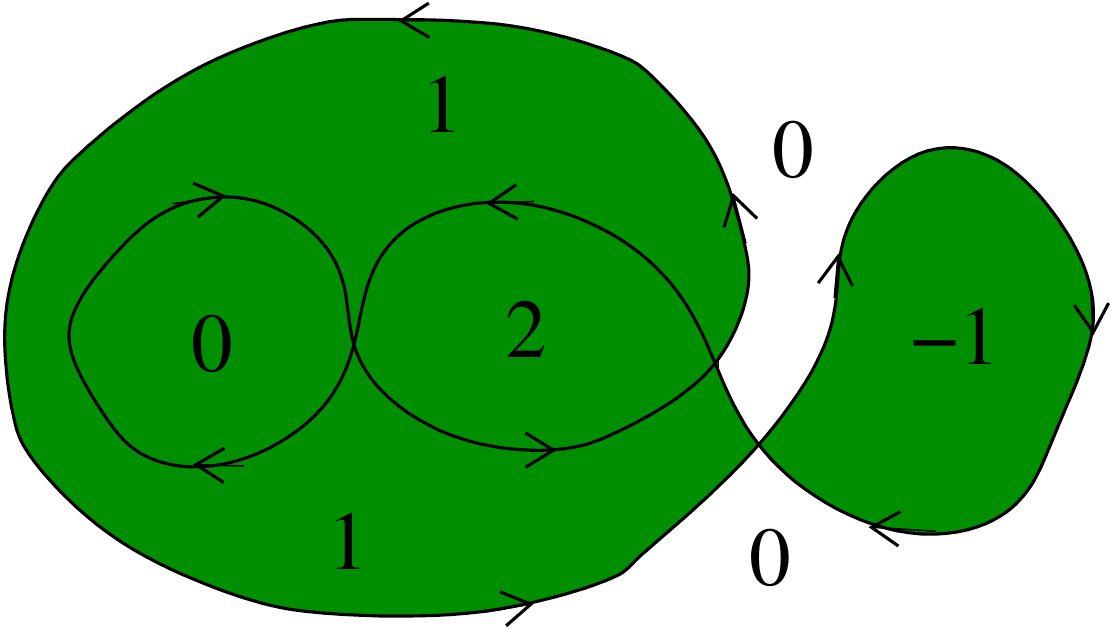}
\vspace{.3in}
\caption{A stylized Brownian loop. The numbers indicate the winding numbers of the loop that contribute additively to $N_{w}$, while the green shaded region is the interior of the loop (the set of points disconnected from infinity by the loop) that contributes $\pm 1$ (where the sign is a Boolean variable assigned randomly to each loop) to the layering
number $N_\ell$.}
\label{winds}
\end{figure}

In this paper we will define and compute certain statistical correlation functions that characterize aspects of the BLS distribution.
We will focus on two types of information:  the number of distinct loops that surround (or ``cover'' if one thinks of the loops as
being filled in) a given point or set of points, and the net number of windings of all the loops around a given point or points
(see \figref{winds}). In both cases, we find results consistent with the correlation functions of primary operators in a conformal
field theory.

\begin{figure}
\begin{center}$
\begin{array}{c c }
	\includegraphics[angle=0, width=0.5\textwidth]{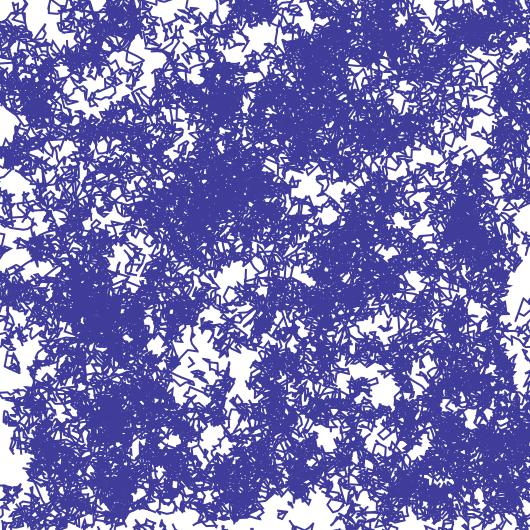} & \includegraphics[angle=0, width=0.5\textwidth]{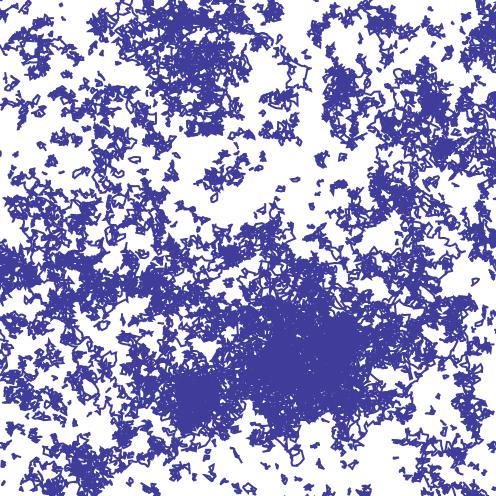} \\
\end{array}$
\end{center}
\caption{The  Brownian loop soup for intensities $\lambda = 2.5$ (left panel) and $\lambda = .5$ (right panel).  The plots were made with both a ``UV'' cutoff on short timelengths and an ``IR'' cutoff on long timelengths.  The true BLS is a scale-invariant fractal that covers every point with probability 1.}  
\label{fig:difference}
\end{figure}

\subsection{Motivation}

We have several motivations for this work.  In \cite{Freivogel:2009rf} one of us considered a similar model, where instead of Brownian loops one sprinkles disks.  The 2- and 3-point correlation functions of certain operators in that model  behave like those of a conformal field theory (CFT), and with a novel set of conformal dimensions.   This is of interest because the model of \cite{Freivogel:2009rf} was derived in \cite{Freivogel:2009it} as an approximation to the asymptotic distribution of bubble nucleations in theories of eternal inflation, a theory for which there is some reason to believe a CFT dual may exist \cite{Maldacena:1997re, Strominger:2001pn,Susskind:2007pv}. 
However, \cite{Freivogel:2009rf} computed the 4-point function exactly, and it suffered from a deficiency:  it was not smooth as the position of the fourth point crossed the circle connecting the other three.  

The origin of the non-analyticity in \cite{Freivogel:2009rf} is probably the fact that although the disk distribution is invariant under global conformal transformations, it is not locally conformally invariant (since disks do not map to disks).  By contrast the BLS distribution is fully conformally invariant.  Therefore we expect the analogous correlation functions to be better behaved, perhaps defining a healthy CFT that could be related to the physics of de Sitter spacetime and eternal inflation.

Another motivation for considering CFTs related to the BLS is the relation between the BLS and SLE \cite{1999math......4022S}, which in turn is related to a large class of conformally invariant models ranging from percolation \cite{2009arXiv0909.4499S, 2006CMaPh.268....1C,2006math......4487C,newcambook} to the Ising model \cite{2013arXiv1312.0533C}.

If in fact the BLS correlation functions we study arise from or define a CFT, it does not seem to be one that is currently known, and it has several interesting and novel features.  As in \cite{Freivogel:2009rf}, the conformal dimensions of the primary operators are real and positive, but vary continuously and as a \emph{periodic} function of a real parameter $\beta$.  As we will see this periodicity arises because the operators are of the form $e^{i \beta N}$, where $N$ is integer valued.

\subsection{Correlation functions in the Brownian Loop Soup} \label{defsec}

As mentioned above, we will study the correlation functions (\emph{i.e.}~expectation values) of two distinct types of operators in the BLS.

\paragraph{Layering number:}  The first type is  closely related to the operators considered in  \cite{Freivogel:2009rf} (which we  refer to  as the ``disk model'').  For each loop, define the interior as the set of  points inside the outermost edge of the loop (including any isolated ``islands'' that might appear inside due to self-intersections; \emph{cf.}~\figref{winds}).  Points in this set are ``covered'' by the loop.  For a  point $z$, consider an operator $N(z)$ that counts the number of distinct loops that cover the point $z$, so that $N(z)$ is the number of ``layers'' at $z$.

One difficulty arises immediately. Because the BLS is conformally (and therefore scale-) invariant, any given
point of the plane is covered by infinitely many loops with probability one.
Since $N(\cdot) \geq 0$, this means that, for any fixed $z$, $\la N(z) \ra$ diverges with probability one. Precisely the same difficulty arises in the disk model, where it was dealt with by adding another, identical and independent copy of the distribution, and then counting the difference in the number of disks of each type that cover $z$.\footnote{In this disk model  this can be thought of as counting the number of bubble transitions that have affected the point $z$, in a model where the field has a discrete shift symmetry and there is a simple rule for bubble collisions (namely the one discussed in \cite{Easther:2009ft}).}  We will follow a very similar procedure here, assigning to each loop a random Boolean value and then defining the layering operator $N_{\ell}(z) \equiv N_{+}(z)-N_{-}(z)$. 

\paragraph{Winding number:}  The other operator we will discuss is $N_{w}(z)$, which counts the total number of windings of all loops around a point $z$ (\figref{winds}). This makes use of the fact that Brownian loops have an orientation (they grow in a particular direction as the time $t$ increases).  Since the winding number can be positive or negative, 
it is not necessary to include another copy of the distribution or compute a difference of two values.  

A natural physical interpretation of winding in the BLS is as follows.  If each loop represents the configuration of a string at some instant of time in a 2+1 dimensional spacetime, winding number counts the number of units of the flux the string is charged under (see \emph{e.g.} \cite{Kleban:2011cs}).

\bigskip

In both cases we will focus on exponentials of these  operators times imaginary coefficients (correlators of the number operators themselves are afflicted by logarithmic divergences, like massless fields in two dimensions).  Due to the similarity to free-field vertex operators we will denote these as 
$$V_{\beta}(z)\equiv e^{i \beta N(z)},$$
where $N$ can be either a layering or winding number operator.\footnote{A special case of the winding model on a lattice was considered by
Le Jan (see Section~6 of \cite{MR2815763}).}

\bigskip

Our paper is written for a mixed audience of mathematicians and physicists.  In most of the paper we  present rigorous proofs of our results.
Sections \ref{dimexp} and \ref{sec-central-charge}, where we perform ``physics-style'' calculations, are the exceptions.
The appendix, Section \ref{appendix}, is dedicated to two important lemmas about the Brownian loop measure \eqref{brownian-loop-measure}
which are used several times in the rest of the paper. A reader uninterested in our methods may simply read \secref{sum} for a summary of  results,
and \secref{conclusions} for our conclusions.

\section{Summary and results} \label{sum}

Our main results relate to correlation functions (\emph{i.e.}~expectation values of products) of exponentials of the winding and layering operators in the BLS.  Specifically, we establish the following:

\begin{itemize}

\item For both versions in finite domains $D$, correlators of $n \in {\mathbb N}$ exponential operators 
\be \label{cor} 
\left\la \Pi_{j} V_{\beta_{j}}(z_{j}) \right\ra_{\delta,D} = \left\la e^{i \sum_{j} \beta_{j} N(z_{j}) } \right\ra_{\delta,D} 
\ee
exist as long as a short-time cutoff $\delta>0$ on the loops is imposed, and
\begin{equation} \nonumber
\lim_{\delta \to 0} 
\frac{\left\langle \prod_{j=1}^n V_{\beta_j}(z_{j})\right\rangle_{\delta, D}}{\prod_{j=1}^n \delta^{2\Delta(\beta_j)}} \equiv
\phi_D(z_1, \dots, z_n; \beta_1, \dots, \beta_n) \equiv \phi_D (\boldsymbol{z}; \boldsymbol{\beta}) 
\end{equation}
exists and is finite. Moreover, if $D'$ is another finite domain and $f:D \to D'$ is a conformal map such that
$z'_1=f(z_1),\ldots,z'_n=f(z_n)$, then
\begin{equation} \nonumber
\phi_{D'}(\boldsymbol{z'};\boldsymbol{\beta}) =
\prod_{j=1}^n \left| f'(z_j) \right|^{-2\Delta(\beta_j)} \phi_D(\bs{z};\boldsymbol{\beta}) \, ,
\end{equation}
where the $\Delta(\beta)$ is defined below.  This is the behavior expected for a conformal primary operator.

\item For both versions in infinite volume, correlators of $n$ exponential operators 
\be \label{cor} 
\left\la \Pi_{j} V_{\beta_{j}}(z_{j}) \right\ra_{\delta} = \left\la e^{i \sum_{j} \beta_{j} N(z_{j}) } \right\ra_{\delta} 
\ee
vanish. 
However, in the case of the layering model, one can remove the short-time cutoff and still obtain a nontrivial
limit by imposing the following ``charge conservation'' condition, satisfied mod $2 \pi$,
\be \label{cc}
\sum_{j} \beta_{j} = 2 \pi k, \,  \,  \,  k \in {\mathbb Z}.
\ee

\item The correlators \eqref{cor} of layering operators in the plane are finite and non-zero when \eqref{cc}
is satisfied, so long as the loop soup is cut off at short times $\delta$ (no long-time cutoff is necessary). 

\item In the case of 2 points, assuming \eqref{cc}, the $\delta \to 0$ limit of the renormalized layering operator
correlators in the plane \eqref{cor} can be explicitly computed up to an overall multiplicative constant.
The result is
\be \nonumber
 \phi_{\mathbb C}(z_1,z_2;\beta_1,\beta_2) = C_{2} \left| \left( {1 \over z_1
     - z_2} \right)^{ \Delta(\beta_1)+\Delta(\beta_2)}  \right|^{2},
 \ee
where $C_{2}$ is a constant (Theorem~\ref{thm-two-point-function}).

\item The $\delta \to 0$ limit of the renormalized 3-point function for the layering model in the plane,
assuming \eqref{cc}, is
\begin{eqnarray*}
\lefteqn{ \phi_{\mathbb C}(z_1,z_2,z_3;\beta_1,\beta_2,\beta_3) =} \\
& & C_{3} \left| \left( {1 \over |z_1 - z_2|} \right)^{\Delta(\beta_1) + \Delta(\beta_2) - \Delta(\beta_3)}  \left( {1 \over |z_1 - z_3|} \right)^{\Delta(\beta_1) + \Delta(\beta_3) - \Delta(\beta_2)} \left( {1 \over |z_2 - z_3|} \right)^{\Delta(\beta_2) + \Delta(\beta_3) - \Delta(\beta_1)} \right|^2 ,
\end{eqnarray*}
where $C_{3}$ is a constant (Theorem~\ref{3-point-function}).

\item The conformal dimensions $\Delta(\beta)$ differ for the two types of operators.  For the layering number,
$$
\Delta_{\ell}(\beta) = {\lambda \over 10} (1-\cos \beta).
$$
For the winding number, 
$$
\Delta_{w}(\beta) = \lambda \beta (2 \pi-\beta)/8 \pi^2,
$$
where this formula is valid for $0\leq \beta < 2 \pi$, and $\Delta_{w}(\beta)$ is periodic under $\beta \to \beta+2 \pi$.  

\end{itemize}

\paragraph{Open questions:}
Our results leave a number of  questions to be answered.  
\begin{itemize}

\item We have not  determined whether the correlators of the winding operators converge in the infinite plane.  The missing piece is the winding number analog of the fact that the BLS is thin \cite{2010arXiv1009.4782N} (see \secref{2-point-function-layering-model}).  It is plausible that these correlators do not in fact converge without an explicit long-distance regulator even when charge conservation \eqref{cc} is satisfied.

\item We have not established that the $n$-point correlators in either model are analytic functions of the $z_{i}$, although we expect this is the case.

\item We expect that these correlators define some kind of conformal field theory.  What conformal field theory is it?

\item The disk model of \cite{Freivogel:2009rf} could be thought of as the late-time distribution of bubbles produced by a first-order phase transition in de Sitter spacetime.  Is there an analogous physical interpretation of the BLS?

\end{itemize}

\section{Correlators of the layering and winding operators}

\subsection{Correlators of layering operator}

As discussed in  \secref{defsec}, Model $1$ is  defined by randomly assigning a Boolean variable to each loop in the Brownian loop soup. 
Alternatively, one can think of this as two independent Brownian loop soups, each with a Poisson distribution $P_{\lambda_{+(-)}, \mu^{loop}}$ with
intensity measure $\lambda_{+(-)}\mu^{loop}$, where we take $\lambda_+ = \lambda_- = \lambda/2$.
(This follows from the fact that the collection of all loops from a BLS of intensity $\lambda_{+}$ and an independent
one of intensity $\lambda_{-}$ is distributed like a BLS with intensity $\lambda_{+} + \lambda_{-}$.)

Denote by $N_{+(-)}(z)$ the number of loops $\gamma$ in the first (respectively, second) class such that the
point $z \in \mathbb C$ is separated from infinity by the image of $\gamma$ in $\mathbb C$.  If $\bar \gamma$ is the ``filled-in"
loop $\gamma$, then this condition becomes $z \in \bar \gamma$, or $z$ is covered by $\gamma$.   We are interested in the layering
field $N_{\ell}$, with $N_{\ell}(z) = N_{+}(z)-N_{-}(z)$. This is purely formal as  both $N_{+(-)}(z)$ are infinite with probability one
for any $z$. They are  infinite for two reasons: both because there are infinitely many large loops surrounding $z$ (infrared, or IR, divergence),
and because there are infinitely many small loops around $z$ (ultraviolet, or UV, divergence). 

We will consider correlators of the exponential operator $V_{\beta} = e^{i \beta N_{\ell}(z)}$, and show that there are
choices of $\beta$ that remove the IR divergence and a normalization which removes the UV divergence. Specifically, we are interested in 
the correlators $V_{\boldsymbol{\beta}}(z_1, \dots, z_n) \equiv \prod_{j=1}^n V_{\beta_j}(z_j) = e^{i \sum_{j=1}^n \beta_j N_{\ell}(z_j)}$
and their moments
\be \nonumber
\left\la V_{\boldsymbol{\beta}}(z_1, \dots, z_n) \right\ra \equiv {\mathbb E}_{\lambda}(V_{\boldsymbol{\beta}}(z_1, \dots, z_n))
\ee
where $z_j \in \mathbb C$, $\boldsymbol {\beta}=(\beta_1, \dots, \beta_n) \in \mathbb R^n$, and the expected value
${\mathbb E}_{\lambda}$   is taken
with respect to the distribution $P_{\lambda,\mu^{loop}} \otimes P_{1/2}$, where $P_{\lambda,\mu^{loop}}$ is the Poisson distribution with intensity measure $\lambda\mu^{loop}$ and
$P_{1/2}$ is the Bernoulli distribution with parameter $1/2$ (remember that each loop belongs to one of two classes
with equal probability) or, equivalently, with respect to two independent copies of the Brownian Loop Soup with equal
intensities $\lambda/2$. 

As the field $N_{\ell}$ has both IR and UV divergences, we get a meaningful definition by introducing cutoffs which
restrict the loops to have diameter\footnote{The diameter of a Brownian loop is the largest distance between two points
on its outer boundary. } within some $\delta$ and $R \in \mathbb R^+$, $\delta < R$: let
$\mu^{loop}_{\delta, R}(\cdot)=\mu^{loop}(\cdot \cap \{\gamma:\delta \leq \diam(\gamma) < R\})$
and consider the correlators
\be \nonumber
\left\la V_{\boldsymbol{\beta}}(z_1, \dots, z_n) \right\ra_{\delta, R} \equiv {\mathbb E}_{\lambda,\delta,R}\left(e^{i \sum_{j=1}^n \beta_j N_{\ell}(z_j)}\right),
\ee
where the expectation ${\mathbb E}_{\lambda,\delta,R}$ is 
as the ${\mathbb E}_{\lambda}$ above  with $\mu^{loop}$ replaced by 
$\mu^{loop}_{\delta,R}$.

\subsection{The $1$-point function in the layering model}
In this section we explicitly compute the $1$-point function in the presence of IR and UV cutoffs.
Replacing the area of a filled Brownian loop of time length $1$ with that of a disk of radius $1$,
the result reproduces the $1$-point function in the disk model \cite{Freivogel:2009rf}.
\begin{lem}
For all  $z \in \mathbb C$, we have that
\be \nonumber
\left\la V_{\beta}(z) \right\ra_{\delta, R} = \left(\frac{R}{\delta}\right)^{- \frac{\lambda}{5}(1-\cos\beta)}.
\ee
\end{lem}

\noindent{\bf Proof.}
 In this and other similar results that follow, we first compute 
$\la V_{\beta}(z) \ra_{\delta, R} = {\mathbb E}_{\lambda,\delta, R}\left(e^{i \beta N_{\ell}(z)}\right)$
in terms of  probabilities such as $\alpha_{z, \delta, R}= \mu_{\delta, R} (\gamma: z \in \bar \gamma)$
(see \eqref{MomFromAlpha}). 
Such calculations can be performed by using two independent copies of the BLS, with
random distributions of loops  indicated by $M_1$ and $M_2$ respectively; then writing
$N_{\ell}(z)= \int \mathbbm{1}_{ \bar \gamma \ni z} (d M_1-dM_2) $,
 where $ \mathbbm{1}_A$ is the indicator function of $A$; next,
 observing that the properties of the Poisson distribution imply that
 $ {\mathbb E}_{\lambda,\delta, R}\left(e^{i \beta \int \mathbbm{1}_{ \bar \gamma \ni z} (d M_1-dM_2)}\right)
 = e^{- \int(1-e^{i \mathbbm{1}_{ \bar \gamma \ni z} })d(\lambda \mu^{loop})} $;
 and, finally, computing the integral in this last expression. 
 We now give a  detailed proof based on the representation with randomly colored loops.
 
 Recall that with
 IR and UV cutoffs in place, the field $N_{\ell}(z)$ can be realized as follows. Let $\eta$ be a realization of loops,
and let $\{X_{\gamma}\}_{\gamma \in \eta}$ be a collection of independent Bernoulli symmetric random variables
taking values in $\{-1,1\}$. The quantity
\be \nonumber
N_\ell(z) = \sum_{\gamma \in \eta, z \in \bar \gamma, \delta \leq \diam(\gamma) <R} X_{\gamma}
\equiv {\sum}^* X_{\gamma}
\ee
is finite $P_{\lambda, \mu^{loop}}$ almost surely, since 
$\mu^{loop}\{\gamma: z \in \bar \gamma, \delta \leq \diam(\gamma) <R\}
=\mu_{\delta, R}\{\gamma: z \in \bar \gamma \}  < \infty$
(see \cite{2005math.....11605W}). Now,
\begin{eqnarray*}
\la V_{\beta}(z) \ra_{\delta, R} &=& {\mathbb E}_{\lambda,\delta, R}\left(e^{i \beta N_{\ell}(z)}\right) \\
&=& \sum_{k=0}^{\infty} {\mathbb E}_{\lambda,\delta, R}\left(e^{i \beta N_{\ell}(z)} | \mathcal L_k \right) 
P_{\lambda, \mu_{\delta, R}}(\mathcal L_k),
\end{eqnarray*}
where  $\mathcal L_k= \{\eta: |\{\gamma \in \eta: z \in \bar \gamma,\delta \leq \diam(\gamma) <R\}|=k \}$.
If $X$ denotes a $(\pm 1)$-valued symmetric random variable,
\be \nonumber
{\mathbb E}_{\lambda,\delta, R}\left(e^{i \beta \sum^* X_{\gamma}} | \mathcal L_k \right) =
\left(E\left(e^{i \beta X}\right)\right)^k = (\cos\beta)^k .
\ee

Therefore, for $\alpha_{z, \delta, R}= \mu_{\delta, R} (\gamma: z \in \bar \gamma)$, we have that
\begin{eqnarray}\label{MomFromAlpha}
\la V_{\beta}(z) \ra_{\delta, R} &=& \sum_{k=0}^{\infty} (\cos\beta)^k
\frac{(\lambda \alpha_{z, \delta, R})^k}{k!} e^{-\lambda \alpha_{z, \delta, R}}
 \\
&=& e^{-\lambda \alpha_{z, \delta, R}(1-\cos\beta)}.
\end{eqnarray}
Moreover, by Lemma \ref {FirstLemma} in \secref{appendix},
\be \nonumber
\alpha_{z, \delta, R} = \frac{1}{5} \log\frac{R}{\delta} ,
\ee
which implies
\be \nonumber
\la V_{\beta}(z) \ra_{\delta, R} = \left(\frac{R}{\delta}\right)^{- \frac{\lambda}{5}(1-\cos(\beta))} ,
\ee
as claimed. \fbox{} \\

\subsection{The winding operator} \label{Sec:WindingOperator}

To define the second model, let $N_{w}(z)$ denote the total winding number  about the 
point $z$ of all
loops in a Brownian Loop Soup; as for the layering operators, this is 
a formal definition as
$N_{w}(z)$ might have divergences. Consider again the 
correlators  $V_{\boldsymbol{\beta}}(z_1, \dots, z_n) = e^{i \sum_{j=1}^n \beta_j N_{w}(z_j)}$
and their moments
$
\la V_{\boldsymbol{\beta}}(z_1, \dots, z_n) \ra = E(V_{\boldsymbol{\beta}}(z_1, \dots, z_n))
$
where $z_j \in \mathbb C$, $\boldsymbol {\beta}=(\beta_1, \dots, \beta_n) \in \mathbb R^n$, and the expected
value is taken with respect to the BLS distribution. Denoting by $P_{\lambda, \mu^{loop}}$ the Poisson distribution
with intensity measure $\lambda\mu^{loop}$, and restricting the loops to have diameter between some
$\delta$ and $R \in \mathbb R^+$, with $\delta < R$, we let
$\mu^{loop}_{\delta, R}(\cdot)=\mu^{loop}(\cdot \cap \{\gamma:\delta \leq \diam(\gamma) < R\})$
and consider the correlators
\be \nonumber
\la V_{\boldsymbol{\beta}}(z_1, \dots, z_n) \ra_{\delta, R}= {\mathbb E}_{\lambda,\delta, R}\left(e^{i \sum_{j=1}^n \beta_j N_{w}(z_j)}\right).
\ee

We now explicitly compute the $1$-point function in the presence of IR and UV cutoffs
for the winding model.

\begin{lem}
For all  $z \in \mathbb C$, we have that
\be \label{winddim}
\la V_{\beta}(z) \ra_{\delta, R} = \left(\frac{R}{\delta}\right)^{- \lambda \frac{\beta(2 \pi - \beta)}{4 \pi^2} },
\ee
where the formula is valid for $\beta \in [0,2\pi)$, and for $\beta \not\in [0,2\pi)$, in the right hand side,
$\beta$ should be replaced by $(\beta \mod 2\pi)$.
\end{lem}

\noindent{\bf Proof.}
For a point $z$ and a loop $\gamma$, let
$\theta_{\gamma}(z)$ indicate the winding number of $\gamma$ around $z$.
Moreover, for $\boldsymbol{k}  \in (\mathbb N \cup \{0\})^{\mathbb N}$   let 
\be \nonumber
\mathcal L_{\boldsymbol{k} }= \{\eta: |\{\gamma \in \eta: 
z \in \bar \gamma,\delta \leq {\rm diam}(\gamma) <R,
|\theta_{\gamma}(z)| = m \}|=k_m \text{ for all } m \in \mathbb N \}.
\ee
If a loop $\gamma$ has $|\theta_{\gamma}(z)| = m $, then the winding number
$\theta_{\gamma}(z)$ is $\pm m$ with equal probability under $P_{\lambda, \mu^{loop}}$.
Finally, using Lemma~\ref{SecondLemma} from \secref{appendix}, for $m \geq 1$ we have
\begin{eqnarray*}
\alpha_{z, \delta, R,m} & \equiv &  \mu^{loop}_{\delta, R}(\gamma: z \in \bar {\gamma}, |\theta_{\gamma}(z)| = m) \\
&=& \mu^{loop}(\gamma: z \in \bar \gamma, \delta \leq {\rm diam}(\gamma) < R, |\theta_{\gamma}(z)| = m) \\
&=& \frac{1}{\pi^2 m^2}\log\frac{R}{\delta} .
\end{eqnarray*}

For all $\boldsymbol{k}  \in (\mathbb N \cup \{0\})^{\mathbb N}$ we have 
\be \nonumber
P_{\lambda, \mu_{\delta, R}}( \mathcal L_{\boldsymbol{k} })
= \prod_{m=1}^{\infty} 
\frac{(\lambda \alpha_{z, \delta, R,m})^{k_m}}{{k_m}!} e^{-\lambda \alpha_{z, \delta, R,m}} ,
\ee
as for different $m$'s the sets of loops with those winding numbers are disjoint.
Hence with IR and UV cutoffs in place, denoting by $E_{\lambda,\delta,R}$ the expectation
with respect to the Poisson distribution $P_{\lambda,\mu^{loop}_{\delta,R}}$ with intensity measure
$\mu^{loop}_{\delta,R}$, we have, for all $z$,
\begin{eqnarray*}
\left\la V_{\beta}(z) \right\ra_{\delta, R} &=& E_{\lambda,\delta,R}\left(e^{i \beta N_{w}(z)}\right) \\
&=& \sum_{\boldsymbol{k}  \in (\mathbb N \cup \{0\})^{\mathbb N}} E_{\lambda,\delta,R}
\left(e^{i \beta N_{w}(z)} | \mathcal L_{\boldsymbol{k} }\right) 
P_{\lambda, \mu_{\delta, R}}(\mathcal L_{\boldsymbol{k} })\\
&=&  \sum_{\boldsymbol{k}  \in (\mathbb N \cup \{0\})^{\mathbb N}}
\prod_{m=1}^{\infty} (\cos(m \beta))^{k_m}
\frac{(\lambda \alpha_{z, \delta, R,m})^{k_m}}{{k_m}!} e^{-\lambda \alpha_{z, \delta, R,m}}
\\
&=& \prod_{m=1}^{\infty}e^{-\lambda \alpha_{z, \delta, R,m}(1-\cos(m \beta))}
\\
&=&\left(\frac{R}{\delta}\right)^{-\lambda  \sum_{m=1}^{\infty}\frac{1}{\pi^2 m^2}(1-\cos(m \beta))}
= \left(\frac{R}{\delta}\right)^{- \lambda \frac{\beta(2 \pi - \beta)}{4 \pi^2} } ,
\end{eqnarray*}
where the $\beta$ on the right hand side of the last equality is to be interpreted modulo $2\pi$. \fbox{} \\

This  coincides with the result  \eqref{physwind} computed using physics path integral methods.

\subsection{The $2$-point function in the layering model} \label{2-point-function-layering-model}
We now analyze the $2$-point function when the IR cutoff is removed by the charge conservation condition \eqref{cc}.

\begin{theorem} \label{thm-two-point-function}
If $\beta_{1}+\beta_{2}= 2 k \pi$ with $k \in {\mathbb Z}$, there is a positive constant $C_2 < \infty$ such that, for all $z_1 \neq z_2$,
\be \nonumber
\lim_{R \to \infty} \left\la V_{\beta_1}(z_1) V_{\beta_2}(z_2) \right\ra_{\delta, R}
= C_2 \left(\frac{|z_1-z_2|}{\delta} \right)^{-\frac{\lambda}{5}(2-\cos \beta_{1}- \cos \beta_{2})} .
\ee
As a consequence,
\be \nonumber
\lim_{\delta \to 0} \lim_{R \to \infty}
\frac{\left\la V_{\beta_1}(z_1) V_{\beta_2}(z_2) \right\ra_{\delta, R}}{\delta^{\frac{\lambda}{5}(2-\cos \beta_{1}- \cos \beta_{2})}}
= C_2 |z_1-z_2|^{-\frac{\lambda}{5}(2-\cos \beta_{1}- \cos \beta_{2})} .
\ee
\end{theorem}

\noindent{\bf Proof.}
Letting $d \equiv |z_{1}-z_{1}|$, for given $\beta_1$ and $\beta_2$, and $d \geq \delta$, we have that
\begin{eqnarray*}
\left\la V_{\beta_1}(z_1) V_{\beta_2}(z_2) \right\ra_{\delta, R} 
& = & \left\la e^{i (\beta_1 N_{\ell}(z_1)+ \beta_2 N_{\ell}(z_2))} \right\ra_{\delta, R} \\
& = & \left\la e^{i (\beta_1 + \beta_2 ) N_{12}} \right\ra_{d, R}
\left\la e^{i \beta_1  N_{1}} \right\ra_{\delta, R} \left\la e^{i  \beta_2  N_{2}} \right\ra_{\delta, R} \, , 
\end{eqnarray*}
where $N_{12}$ is the number of loops that cover both
$z_1$ and $z_2$,  and $N_1$ ($N_2$) is the number of loops that cover 
$z_1$ but not  $z_2$ ($z_2$ but not  $z_1$, resp.). The two-point function
factorizes because the sets of loops contributing to $N_{1 2}$,
$N_1$ and $N_2$ are disjoint; the $\delta$ is replaced by $d$ in the first factor
in the second line because a loop covering both $z_1$ and $z_2$ must have
diameter at least $d$.

As in the $1$-point function calculation, we can write
\be \nonumber
\left\la e^{i (\beta_1 + \beta_2 ) N_{12}} \right\ra_{d, R} =
\sum_{n=0}^{\infty} \left(\cos(\beta_1 + \beta_2)\right)^n P_{\lambda, \mu^{loop}_{d, R}}(N_{1 2}=n)
= e^{-\lambda \alpha_{d, R}(z_1,z_2)\left( 1 - \cos(\beta_1 + \beta_2) \right)},
\ee
where $\alpha_{d, R}(z_1,z_2) \equiv \mu^{loop}_{d, R}(\gamma : z_1, z_2 \in \bar \gamma)$.
Similarly, if 
$\alpha_{\delta, R}(z_1,\neg z_2) = \mu^{loop}_{\delta, R}(\gamma : z_1 \in \bar \gamma,
z_2 \notin \bar \gamma)$ and $\alpha_{\delta, R}(\neg z_1,  z_2)$ is correspondingly
defined, then
\be \nonumber
\left\la e^{i \beta_1 N_{1}} \right\ra_{\delta, R} = e^{-\lambda\alpha_{\delta, R}(z_1,\neg z_2)(1-\cos\beta_1)}
\ee
and
\be \nonumber
\left\la e^{i \beta_2 N_{2}} \right\ra_{\delta, R} = e^{-\lambda\alpha_{\delta, R}(\neg z_1,z_2)(1-\cos\beta_2)} \, .
\ee
Combining the three terms we obtain
\begin{eqnarray*}
\lefteqn{
 \left\la e^{i (\beta_1 N_{\ell}(z_1)+ \beta_2 N_{\ell}(z_2))} \right\ra_{\delta, R} } \\
 & = & e^{-\lambda\alpha_{d, R}(z_1,z_2)(1-\cos(\beta_1 + \beta_2))} \,
 e^{-\lambda\alpha_{\delta, R}(z_1,\neg z_2)(1-\cos\beta_1)} \, e^{-\lambda\alpha_{\delta, R}(\neg z_1,z_2)(1-\cos\beta_2)} \, .
 \end{eqnarray*}
It is easy to see that $\lim_{R \rightarrow \infty}\alpha_{d, R}(z_1,z_2) = \infty$.
(This follows from the scale invariance of $\mu^{loop}$ by considering an increasing -- in size -- sequence of disjoint,
concentric annuli around $z_{1}$ and $z_{2}$ that are scaled versions of each other.) Hence, in order to remove the
IR cutoff, we must impose \eqref{cc}
and set $\beta_{1}+\beta_{2}= 2 k \pi$, so that  $1-\cos(\beta_1 + \beta_2)=0$.  
 
Assuming that $\beta_{1}+\beta_{2}= 2 k \pi$, we are left with
 \be \label{2pf-charge-cons}
  \left\la e^{i (\beta_1 N_{\ell}(z_1)+ \beta_2 N_{\ell}(z_2))} \right\ra_{\delta, R}
 =e^{
 -\lambda\alpha_{\delta, R}(z_1,\neg z_2)(1-\cos\beta_1)-\lambda\alpha_{\delta, R}(\neg z_1,z_2)(1-\cos\beta_2)} .
 \ee
To remove the infrared cutoff, we use the fact that the loop is \emph{thin}: If $z_1 \neq z_2$,
$\mu^{loop}(\gamma:z_1 \in \bar \gamma, z_2 \notin \bar \gamma, \diam(\gamma) \geq \delta)<\infty$
for any $\delta>0$ (see \cite{2010arXiv1009.4782N}, Lemma 4).
By the obvious monotonicity of $\alpha_{ \delta, R} (z_1,\neg z_2)$ in $R$, this implies that
\begin{equation*}
\lim_{R \rightarrow \infty} \alpha_{\delta, R}(z_1,\neg z_2) = \mu^{loop}\{\gamma \in \eta: z_{1} \in \bar \gamma, 
 z_{2} \notin  \bar \gamma, {\rm diam}(\gamma) \geq \delta \} \equiv \alpha_{\delta}(z_1,\neg z_2) .
\end{equation*}
By scale, rotation and translation invariance of the Brownian loop measure $\mu^{loop}$,
$\alpha_{\delta}(z_1,\neg z_2) $ can only depend on the ratio $x=d/\delta$, so we can introduce the notation
$
\alpha(x) \equiv \alpha_{\delta}(z_1,\neg z_2).
$
The function $\alpha$ has the following properties, which are also immediate consequences of the scale,
rotation and translation invariance of the Brownian loop measure.
 \begin{itemize}
 \item $ \alpha(x) =\alpha_{\delta}(0,\neg z)$ for any $z$ such that $|z|=d$.
 \item For $\sigma \geq 1$, if $\delta < d$, letting $\alpha_{\delta, R}(z) \equiv \alpha_{z, \delta, R}=\mu_{\delta, R}(\gamma: z \in \bar \gamma)$,
 \begin{eqnarray} \label{alpha-sigma}
  \alpha( \sigma x) &=&\alpha_{\delta}(0,\neg \sigma z) \nonumber \\
  &=& \alpha_{\sigma \delta}(0,\neg \sigma z) + \alpha_{\delta, \sigma \delta}(0,\neg \sigma z) \nonumber \\
  &=&   \alpha(  x) +  \alpha_{\delta, \sigma \delta}(0)
   = \alpha(  x) +  \alpha_{1, \sigma }(0).
  \end{eqnarray}
 \end{itemize}

Now let 
 \be \nonumber
 G(x) \equiv \left\la e^{i (\beta_1 N_{\ell}(z_1)+ \beta_2 N_{\ell}(z_2))} \right\ra_{\delta}
 \equiv \lim_{R \rightarrow \infty}  \left\la e^{i (\beta_1 N_{\ell}(z_1)+ \beta_2 N_{\ell}(z_2))} \right\ra_{\delta, R} \, ;
 \ee
using \eqref{2pf-charge-cons} and the definition of the function $\alpha$, we can write 
 \be \nonumber
 G(x)=e^{- \lambda \alpha(x)(2-\cos \beta_{1}- \cos \beta_{2})} \, .
 \ee
Then, for $\sigma \geq 1$,
\be \nonumber
 G(\sigma x)=e^{- \lambda \alpha_{1,\sigma}(0)(2-\cos \beta_{1}- \cos \beta_{2})} G(x).
 \ee
Using Lemma~\ref{FirstLemma}, we have that
 \be \nonumber
 \alpha_{1,\sigma}(0) = \frac{1}{5} \log \sigma .
 \ee
 It then follows that, for $\sigma \geq 1$,
 \be \label{scaling-eq}
 G(\sigma x)= \sigma^{- \frac{\lambda}{5} (2-\cos \beta_{1}- \cos \beta_{2}) } G(x) .
 \ee

For $0<\sigma<1$, \eqref{alpha-sigma} implies
\be \nonumber
\alpha(\sigma x) = \alpha(x) - \alpha_{1,1/\sigma }(0).
\ee
But since
\be \nonumber
\alpha_{1,1/\sigma}(0) = -\frac{1}{5} \log \sigma ,
\ee
equation~\eqref{scaling-eq} is unchanged when $0<\sigma<1$.

The fact that \eqref{scaling-eq} is valid for all $\sigma>0$ immediately implies that
 \be \nonumber
G(x) = C_2 x^{-\frac{\lambda}{5}(2-\cos \beta_{1}- \cos \beta_{2})}
\ee
for some constant $C_2 >0$. \fbox{}

\section{Conformal covariance of the $n$-point functions}
We now analyze the $n$-point functions for general $n \geq 1$ and their conformal invariance properties.
In bounded domains $D \subset {\mathbb C}$, we show, for both models, how to remove the UV cutoff $\delta>0$
by dividing by $\delta^{ 2 \sum_{j=1}^n \Delta_{j}}$, with the appropriate $\Delta_j$'s. We also show that this procedure leads
to conformally covariant functions of the domain $D$. The scaling with $\delta$ originates from the fact that loops with diameter
less than $\delta$ can only wind around a single point in the limit $\delta \to 0$, and so for these small loops the $n$-point function
reduces to the product of 1-point functions.

In Section~\ref{n-point-plane}, we deal with the layering model in the full plane, $\mathbb C$, and show that, together
with the UV cutoff $\delta>0$, we can also remove the IR cutoff $R<\infty$, provided we impose the condition
$\sum_{j=1}^n \beta_j \in 2 \pi \Z$ (\emph{cf.} \eqref{cc}).  We refer to this condition as  ``charge conservation''
because---apart from the periodicity---it is reminiscent of momentum or charge conservation for the vertex operators
of the free boson.

Just as for the two-point function, in the layering model the IR convergence (given ``charge conservation'') is due to the finiteness of the total mass of the loops which
cover some points but not others; this is basically the property that the soup of outer boundaries of a Brownian loop soup is \emph{thin}
in the language of Nacu and Werner \cite{2010arXiv1009.4782N}. We did not prove the analogous finiteness for the winding model,
so in that case we are not able to prove that the IR cutoff can be removed.

\subsection{The layering model in finite domains} \label{n-point-finite-1}

\medskip\noindent

\medskip\noindent

In the theorem below, we let
$\left\langle \prod_{j=1}^n V_{\beta_j}(z_j) \right\rangle_{\delta,D} =
{\mathbb E}_{\lambda,\delta,D}\left(\prod_{j=1}^n e^{i \beta_j N_\ell(z_j)}\right)$ denote
the expectation of the product $\prod_{j=1}^n e^{i \beta_j N_\ell(z_j)}$ with respect to a loop soup in $D$
with intensity $\lambda>0$ containing only loops of diameter at least $\delta>0$, that is, with respect to the
distribution $P_{\lambda,\mu^{loop}_{\delta,D}} \otimes P_{1/2}$, where $P_{\lambda,\mu^{loop}_{\delta,D}}$
is the Poisson distribution with intensity measure
$\mu^{loop}_{\delta,D}
=\mu^{loop}_D \mathbbm{1}_{\{\diam(\gamma)\geq\delta\}}
=\mu^{loop} \mathbbm{1}_{\{\gamma \subset D, \diam(\gamma)\geq\delta\}}$
and $P_{1/2}$ is the Bernoulli distribution with parameter $1/2$ (remember that each loop belongs to one of two classes
with equal probability).
\begin{theorem} \label{thm-bounded-domains}
If $n \in {\mathbb N}$, $D \subset {\mathbb C}$ is bounded  and
$\boldsymbol{\beta}=(\beta_1,\ldots,\beta_n)$, then 
\begin{equation} \nonumber
\lim_{\delta \to 0} 
\frac{\left\langle \prod_{j=1}^n V_{\beta_j}(z_j)\right\rangle_{\delta, D}}{\delta^{ \lambda \frac{1}{5}\sum_{j=1}^n (1-\cos\beta_j)}} \equiv \phi_D(z_1, \dots, z_n;\boldsymbol{\beta}) 
\end{equation}
exists and is finite and real. Moreover, if $D'$ is another bounded subset of $\mathbb C$ and $f:D \to D'$ is a conformal map such that
$z'_1=f(z_1),\ldots,z'_n=f(z_n)$, then
\begin{equation} \nonumber
\phi_{D'}(z'_1,\ldots,z'_n;\boldsymbol{\beta}) = \prod_{j=1}^n |f'(z_j)|^{-\frac{\lambda}{5}(1-\cos\beta_j)} \phi_D(z_1, \dots, z_n;\boldsymbol{\beta}) \, .
\end{equation}
\end{theorem}

The proof of the theorem will make use of the following lemma, where $B_{\delta}(z)$ denotes the disc of radius $\delta$ centered at $z$,
$\bar\gamma$ denotes the complement of the unique unbounded component of ${\mathbb C} \setminus \gamma$, and where $o(1)$ denotes
a quantity ``smaller than $O(1)$'', i.e. that tends to zero as $\delta \to 0$.
\begin{lemma} \label{lemma}
Let $D,D' \subset\mathbb C$ and let $f:D \to D'$ be a conformal map. For $n\geq1$, assume that $z_1,\ldots,z_n \in D$ are distinct
and that $z'_1=f(z_1),\ldots,z'_n=f(z_n)$, and let $s_j=|f'(z_j)|$ for $j=1, \ldots, n$ . Then we have that, for each $j=1,\ldots,n$,
\begin{eqnarray*} \label{eq-lemma}
\lefteqn{\mu^{loop}_{D'}(\gamma: z'_j \in \bar \gamma, \bar\gamma \not\subset f(B_{\delta}(z_j)), z'_k \notin \bar\gamma \; \forall k \neq j) } \\
& - & \mu^{loop}_{D'}(\gamma: z'_j \in \bar \gamma, \bar\gamma \not\subset B_{s_j \delta}(z'_j), z'_k \notin \bar\gamma \; \forall k \neq j) = o(1) \text{ as } \delta \to 0 .
\end{eqnarray*}
\end{lemma}

\noindent{\bf Proof.}
Let $B_{in}(z'_j)$ denote the largest (open) disc centered at $z'_j$ contained inside $f(B_{\delta}(z_j)) \cap B_{s_j \delta}(z'_j)$,
and $B_{out}(z'_j)$ denote the smallest disc centered at $z'_j$ containing $f(B_{\delta}(z_j)) \cup B_{s_j \delta}(z'_j)$. A moment
of thought reveals that, for $\delta$ sufficiently small,
\begin{eqnarray} \label{upper-bound}
\lefteqn{|\mu^{loop}_{D'}(\gamma: z'_j \in \bar \gamma, \bar\gamma \not\subset f(B_{\delta}(z_j)), z'_k \notin \bar\gamma \; \forall k \neq j) } \nonumber \\
& - & \mu^{loop}_{D'}(\gamma: z'_j \in \bar \gamma, \bar\gamma \not\subset B_{s_j \delta}(z'_j), z'_k \notin \bar\gamma \; \forall k \neq j)| \nonumber \\
& = & \mu^{loop}_{D'}(\gamma: z'_j \in \bar\gamma, \bar\gamma \not\subset f(B_{\delta}(z_j)) \cap s_j B_{\delta}(z'_j), \bar\gamma \subset f(B_{\delta}(z_j)) \cup s_j B_{\delta}(z'_j)
 , z'_k \notin \bar\gamma \; \forall k \neq j) \nonumber \\
& \leq & \mu^{loop}_{D'}(\gamma: z'_j \in \bar\gamma, \bar\gamma \not\subset B_{in}(z'_j), \bar\gamma \subset B_{out}(z'_j)) \nonumber \\
& = &  c \log\frac{\diam(B_{out}(z'_j))}{\diam(B_{in}(z'_j))} \, ,
\end{eqnarray}
where $c<\infty$ is a positive constant and the last equality follows from Proposition~3 of \cite{2005math.....11605W}.
Note that, when $D' = {\mathbb C}$, the quantities above involving $\mu^{loop}_{\mathbb C}$ are bounded because of the fact that
the Brownian loop soup is \emph{thin} \cite{2010arXiv1009.4782N}.

Since $f$ is analytic, for every $w \in \partial B_{\delta}(z_j)$, we have that
\begin{equation} \nonumber
|f(w)-z'_j| = 
s_j \delta + O(\delta^2) \, ,
\end{equation}
which implies that
\begin{equation} \nonumber
\lim_{\delta \to 0} \frac{\diam(B_{out}(z'_j))}{\diam(B_{in}(z'_j))}
= 1 \, .
\end{equation}
In view of \eqref{upper-bound}, this concludes the proof of the lemma. \fbox{} \\

\noindent{\bf Proof of Theorem \ref{thm-bounded-domains}.}
We first show that the limit is finite. 
Using the notation of the previous section, we let $\eta$ denote a realization of loops and
$\{X_{\gamma}\}_{\gamma \in \eta}$ a collection of independent Bernoulli symmetric random variables taking values in $\{-1,1\}$.
Moreover, let $[n] \equiv \{1,\dots,n\}$, let $\mathcal K$ denote the space of assignments of a nonnegative integer to each nonempty
subset $S$ of $\{z_1,\ldots,z_n\}$, and for $S \subset \{z_1,\ldots,z_n\}$, let $I_{S} \subset [n]$ be the set of indices such that
$k \in I_S$ if and only if $z_k \in S$. We have that
\begin{eqnarray*}
\left\langle \prod_{j=1}^n V_{\beta_j}(z_j) \right\rangle_{\delta,D}
&=& {\mathbb E}_{\lambda,\delta,D}\left(e^{i \sum_{j=1}^n \beta_j N_{\ell}(z_j)}\right)  \\
&=& \sum_{\boldsymbol{k} \in \mathcal K} {\mathbb E}_{\lambda,\delta,D}\left(e^{i \sum_{j=1}^n \beta_j N_{\ell}(z_j)} | \mathcal L_{\boldsymbol{k} }\right) 
P_{\lambda, \mu^{loop}_{\delta, D}}(\mathcal L_{\boldsymbol{k} })
\end{eqnarray*}
where  $\mathcal L_{\boldsymbol{k} }= \{\eta: \forall S \subset \{z_1,\ldots,z_n\}, S \neq \emptyset,
|\{\gamma \in \eta: z_j \in \bar \gamma \quad \forall j \in I_{S}, z_j \not \in \bar \gamma \quad \forall j \not \in I_{S} \}|=\boldsymbol{k}(S) \}$.
With probability one with respect to $P_{\lambda, \mu_{\delta, D}}$, we have that for each $j=1,\ldots,n$, 
\be \nonumber
N_{\ell}(z_j)= \sum_{\gamma: z_j \in \bar \gamma, \diam(\gamma) \geq \delta} X_{\gamma}
= \sum_{S \subset \{ z_{1},\ldots,z_{n} \}: z_j \in S} \, \sum_{\gamma: S \subset \bar \gamma, S^c \subset \bar\gamma^c, 
 \diam(\gamma) \geq \delta} X_{\gamma} .
\ee

With the notation $\sum^{S} \equiv \sum_{\gamma: S \subset \bar \gamma, S^c \subset \bar\gamma^c, \diam(\gamma) \geq \delta}$,
and letting $X$ denote a $(\pm 1)$-value symmetric random variable, we have that
 \begin{eqnarray*}
 {\mathbb E}_{\lambda,\delta, R}\left(e^{i \sum_{j=1}^n \beta_j N_{\ell}(z_j)} | \mathcal L_{\boldsymbol{k} } \right) 
 &=&
 {\mathbb E}_{\lambda,\delta,R}\left(e^{i \sum_{j=1}^n \beta_j \sum_{S \subset \{ z_{1},\ldots,z_{n} \}: z_j \in S} 
 \sum^{S} X_{\gamma} } | \mathcal L_{\boldsymbol{k} } \right) \\
 &=&
 {\mathbb E}_{\lambda,\delta, R}\left(e^{i \sum_{S \subset \{ z_{1},\ldots,z_{n} \}} \sum^{S} (\sum_{j \in I_{S}} \beta_j ) X_{\gamma} }
 | \mathcal L_{\boldsymbol{k} }\right)\\
   &=&
 \prod_{S \subset \{z_1,\ldots,z_n\}, S \neq \emptyset} \left(E\left(e^{i (\sum_{j \in I_{S}} \beta_j ) X } \right)\right)^{\boldsymbol{k}(S)} \\
   &=&
 \prod_{S \subset \{z_1,\ldots,z_n\}, S \neq \emptyset} \left(\cos\left(\sum_{j \in I_{S}} \beta_j  \right)\right)^{\boldsymbol{k} (S)} .
  \end{eqnarray*}

Next, given $S \subset \{ z_1, \ldots, z_n \}$ with $|S| \geq 2$, let
$\alpha_D(S) \equiv \mu_{D}(\gamma: S \subset \bar \gamma, S^c \subset \bar\gamma^c)$ and
$\alpha_{\delta,D}(z_j) \equiv \mu_{D}(\gamma: \diam(\gamma) \geq \delta, z_j \in \bar \gamma, z_k \notin \bar\gamma \; \forall k \neq j)$.
Furthermore, let $I_S$ be the set of indices such that $k \in I_S$ if and only if
$z_k \in S$. Let  $m = \min_{i,j: i \neq j} |z_i -  z_j| \wedge \min_i \dist(z_i,\partial D)$ and note that, when $\delta<m$, we can write

\begin{eqnarray}
\left\langle \prod_{j=1}^n V_{\beta_j}(z_j) \right\rangle_{\delta,D}
&=&
\sum_{\boldsymbol{k} \in \mathcal K}  \prod_{S \subset \{z_1,\ldots,z_n \}, |S|>1} \left(\cos\left(\sum_{k \in I_S} \beta_k \right)\right)^{\boldsymbol{k} (S)}
\frac{ (\lambda \alpha_D(S))^{\boldsymbol{k}(S)}}{({\boldsymbol{k} (S)})!} e^{- \lambda \alpha_D(S)} \nonumber \\
& & \prod_{j=1}^n \left( \cos\beta_j \right)^{\boldsymbol{k}(z_j)} \frac{ (\lambda \alpha_{\delta,D}(z_j))^{\boldsymbol{k}(z_j)}}{({\boldsymbol{k}(z_j)})!} e^{- \lambda \alpha_{\delta,D}(z_j)} \nonumber \\
&=& 
\prod_{S \subset \{z_1,\ldots,z_n \}, |S|>1} \exp{\left[-\lambda \alpha_D(S) \left(1-\cos\left(\sum_{k \in I_S} \beta_k \right)\right)\right]} \nonumber \\
& & \prod_{j=1}^n \exp{\left[-\lambda \alpha_{\delta,D}(z_j)(1 - \cos \beta_j) \right]} \, . \label{n-point-func}
\end{eqnarray}

For every $j=1,\ldots,n$, using Lemma~\ref{FirstLemma}, we have that
 \begin{eqnarray*}
  \alpha_{\delta,D}(z_j) &=& \mu^{loop}_{D}(\gamma: \diam(\gamma) \geq \delta, z_j \in \bar \gamma, z_k \notin \bar\gamma \; \forall k \neq j)  \\
  & = & \mu^{loop}_{D}(\gamma: m > \diam(\gamma) \geq \delta, z_j \in \bar \gamma) \\
  & + & \mu^{loop}_{D}(\gamma: \diam(\gamma) \geq m, z_j \in \bar \gamma, z_k \notin \bar\gamma \; \forall k \neq j)\\
  &=&
  \frac{1}{5} \log\frac{m}{\delta} + \alpha_{m,D}(z_j).
 \end{eqnarray*}
Therefore, we obtain
 \begin{eqnarray*}
\lim_{\delta \to 0} 
\frac{\left\langle \prod_{j=1}^n V_{\beta_j}(z_j) \right\rangle_{\delta,D}}{\delta^{ \frac{\lambda}{5}\sum_{j=1}^n (1-\cos(\beta_j))}}
& = & \prod_{S \subset \{z_1,\ldots,z_n \}, |S|>1} \exp{\left[-\lambda \alpha_D(S) \left(1-\cos\left(\sum_{k \in I_S} \beta_k \right)\right)\right]} \\
 & & m^{ -\frac{\lambda}{5}\sum_{j=1}^n (1-\cos\beta_j)} e^{-\sum_{j=1}^n \lambda \alpha_{m,D}(z_j)(1-\cos\beta_j)} \\
& = & m^{ -\frac{\lambda}{5}\sum_{j=1}^n (1-\cos\beta_j)}
\exp{\left[ -\lambda \sum_{j=1}^n \alpha_{m,D}(z_j)(1-\cos\beta_j) \right]} \\
& & \exp{\left[ -\lambda \sum_{S \subset \{z_1,\ldots,z_n \}, |S|>1} \alpha_D(S) \left(1-\cos\left(\sum_{k \in I_S} \beta_k \right)\right) \right]} \\
& \equiv & \phi_D(z_1, \dots, z_n; \boldsymbol{\beta}) \, .
\end{eqnarray*}
This concludes the first part of the proof.

To prove the second part of the theorem, using \eqref{n-point-func}, we write
\begin{eqnarray} \label{n-point-funct}
\left\langle \prod_{j=1}^n V_{\beta_j}(z_j) \right\rangle_{\delta,D} & = & \exp{\left[ -\lambda \sum_{S \subset \{ z_1, \ldots, z_n \}, |S| \geq 2} \alpha_D(S) \left(1-\cos\left(\sum_{k \in I_S} \beta_k \right)\right) \right]} \nonumber \\
                                                              &     & \prod_{j=1}^n \exp{\left[-\lambda \alpha_{\delta,D}(z_j)(1 - \cos \beta_j) \right]} \, .
\end{eqnarray}
For each $S \subset \{ z_1, \ldots, z_n \}$ with $|S| \geq 2$, $\alpha_D(S)$ is invariant under conformal transformations, that is,
if $f:D \to D'$ is a conformal map from $D$ to another bounded domain $D'$, and $S' = \{ z'_1,\ldots,z'_n \}$, where $z'_1=f(z_1),\ldots,z'_n=f(z_n)$,
then $\alpha_{D'}(S') = \alpha_D(S)$. Therefore, the first exponential term in \eqref{n-point-funct} is also invariant under conformal transformations.
This implies that,  for $\delta$ sufficiently small, 
\begin{equation} \nonumber
\frac{\left\langle \prod_{j=1}^n V_{\beta_j}(z_j) \right\rangle_{\delta,D}}{\left\langle \prod_{j=1}^n V_{\beta_j}(z_j') \right\rangle_{\delta,D'}}
= \prod_{j=1}^n \exp{\left\{-\lambda \left[\alpha_{\delta,D}(z_j)-\alpha_{\delta,D'}(z'_j)\right](1 - \cos \beta_j) \right\}} \, . 
\end{equation}

Writing
\begin{eqnarray*}
\alpha_{\delta,D}(z_j) & = & \mu^{loop}_D(\gamma: \diam(\gamma) \geq \delta, z_j \in \bar \gamma, \bar\gamma \subset B_{\delta}(z_j)) \\
& + & \mu^{loop}_D(\gamma: z_j \in \bar \gamma, \bar\gamma \not\subset B_{\delta}(z_j), z_k \notin \bar\gamma \; \forall k \neq j)
\end{eqnarray*}
and noticing that $\mu^{loop}_D(\gamma: \diam(\gamma) \geq \delta, z_j \in \bar \gamma, \bar\gamma \subset B_{\delta}(z_j))
= \mu^{loop}_{D'}(\gamma: \diam(\gamma) \geq \delta, z'_j \in \bar \gamma, \bar\gamma \subset B_{\delta}(z'_j))$ (where we have assumed,
without loss of generality, that $\delta$ is so small that $B_{\delta}(z_j) \subset D$ and $B_{\delta}(z'_j) \subset D'$), we have that
\begin{eqnarray*}
\alpha_{\delta,D}(z_j) - \alpha_{\delta,D'}(z'_j) & = &
\mu^{loop}_D(\gamma: z_j \in \bar \gamma, \bar\gamma \not\subset B_{\delta}(z_j), z_k \notin \bar\gamma \; \forall k \neq j) \\
& - & \mu^{loop}_{D'}(\gamma: z'_j \in \bar \gamma, \bar\gamma \not\subset B_{\delta}(z'_j), z'_k \notin \bar\gamma \; \forall k \neq j) \, .
\end{eqnarray*}
To evaluate this difference, using conformal invariance, we write
\begin{eqnarray*}
\lefteqn{ \mu^{loop}_D(\gamma: z_j \in \bar \gamma, \bar\gamma \not\subset B_{\delta}(z_j), z_k \notin \bar\gamma \; \forall k \neq j) } \\
& = & \mu^{loop}_{D'}(\gamma: z'_j \in \bar \gamma, \bar\gamma \not\subset f(B_{\delta}(z_j)), z'_k \notin \bar\gamma \; \forall k \neq j) \, .
\end{eqnarray*}
Letting $s_j = |f'(z_j)|$ and using \eqref{log-equation} and Lemma \ref{FirstLemma} from the appendix, we can write
\begin{align*}
& \alpha_{\delta,D}(z_j) - \alpha_{\delta,D'}(z'_j) \\
& = \mu^{loop}_{D'}(\gamma: z'_j \in \bar \gamma, \bar\gamma \not\subset f(B_{\delta}(z_j)), z'_k \notin \bar\gamma \; \forall k \neq j)
- \mu^{loop}_{D'}(\gamma: z'_j \in \bar \gamma, \bar\gamma \not\subset B_{\delta}(z'_j), z'_k \notin \bar\gamma \; \forall k \neq j) \\
& = \mu^{loop}_{D'}(\gamma: z'_j \in \bar \gamma, \bar\gamma \not\subset f(B_{\delta}(z_j)), z'_k \notin \bar\gamma \; \forall k \neq j)
- \mu^{loop}_{D'}(\gamma: z'_j \in \bar \gamma, \bar\gamma \not\subset B_{s_j \delta}(z'_j), z'_k \notin \bar\gamma \; \forall k \neq j) \\
& \hspace{.5cm} - [ \mu^{loop}_{D'}(\gamma: z'_j \in \bar \gamma, \bar\gamma \not\subset B_{\delta}(z'_j), z'_k \notin \bar\gamma \; \forall k \neq j)
- \mu^{loop}_{D'}(\gamma: z'_j \in \bar \gamma, \bar\gamma \not\subset B_{s_j \delta}(z'_j), z'_k \notin \bar\gamma \; \forall k \neq j) ] \\
& = \mu^{loop}_{D'}(\gamma: z'_j \in \bar \gamma, \bar\gamma \not\subset f(B_{\delta}(z_j)), z'_k \notin \bar\gamma \; \forall k \neq j)
- \mu^{loop}_{D'}(\gamma: z'_j \in \bar \gamma, \bar\gamma \not\subset B_{s_j \delta}(z'_j), z'_k \notin \bar\gamma \; \forall k \neq j) \\
& \hspace{.5cm} - \frac{1}{5} \log s_j \, .
\end{align*}

Using Lemma \ref{lemma}, we obtain
$\alpha_{\delta,D}(z_j) - \alpha_{\delta,D'}(z'_j) = -\frac{1}{5} \log |f'(z_j)| + o(1)$ as $\delta \to 0$, which gives
\begin{equation} \nonumber
\frac{\left\langle V_{\bf\beta}(z_1,\ldots,z_n) \right\rangle_{\delta,D}}{\left\langle V_{\bf\beta}(z'_1,\ldots,z'_n) \right\rangle_{\delta,D'}}
= e^{-o(1)} \prod_{j=1}^n |f'(z_j)|^{\frac{\lambda}{5}(1-\cos\beta_j)} \text{ as } \delta \to 0.
\end{equation}
Letting $\delta \to 0$ concludes the proof. \fbox{} \\

\subsection{The winding model in finite domains} \label{n-point-finite-2}

As above, let $N_{w}(z)$ denote the total number of windings of all loops of a given soup around $z \in {\mathbb C}$.
We have the following theorem.
\begin{theorem} \label{thm-bounded-domains-winding}
If $n \in {\mathbb N}$, $D \subset {\mathbb C}$ is bounded  and $\boldsymbol{\beta}=(\beta_1,\ldots,\beta_n)$, then
\begin{equation} \nonumber
\lim_{\delta \to 0} 
\frac{\left\langle e^{i \beta_1N_{w}(z_1)} \dots e^{i \beta_n N_{w}(z_n)} \right\rangle_{\delta, D}}{\delta^{ \frac{\lambda}{4\pi^2} \sum_{j=1}^n \beta_j(2\pi-\beta_j)}}
\equiv \psi_D(z_1,\ldots,z_n;\boldsymbol{\beta}) 
\end{equation}
exists and is finite and real. Moreover, if $D'$ is another bounded subset of $\mathbb C$ and $f:D \to D'$ is a conformal map such that
$z'_1=f(z_1),\ldots,z'_n=f(z_n)$, then
\begin{equation} \nonumber
\psi_{D'}(z'_1,\ldots,z'_n;\boldsymbol{\beta}) = \prod_{j=1}^n \left|f'(z_j)\right|^{-\lambda\frac{\beta_j(2\pi-\beta_j)}{4\pi^2}} \psi_D(z_1,\ldots,z_n; \boldsymbol{\beta}) \, ,
\end{equation}
where, in the exponent, the $\beta_j$'s are to be interpreted modulo $2\pi$.
\end{theorem}

\noindent{\bf Proof.}
The proof is analogous to that of Theorem \ref{thm-bounded-domains}. Let
$\alpha_D(S;k_{i_1},\ldots,k_{i_l}) := \mu_D(\gamma: \theta_{\gamma}(z_{i_j}) = k_{i_j} \text{ for each } z_{i_j} \in S \text{ and } S^c \subset \bar\gamma^c)$,
and $\alpha_{\delta,D}(z_j;k) := \mu_{D}(\gamma: z_j \in \bar \gamma, \theta_{\gamma}(z_j) = k, z_k \notin \bar\gamma \; \forall k \neq j)$.
With this notation we can write
\begin{eqnarray*} \label{n-point-funct-winding}
\lefteqn{\left\langle e^{\beta_1N_{w}(z_1)} \ldots e^{\beta_nN_{w}(z_n)} \right\rangle_{\delta,D} } \\
& = & \exp{\left[ -\lambda \sum_{l=2}^n \sum_{\stackrel{S \subset \{ z_1, \ldots, z_n \}}{|S|=l}}
\sum_{k_{i_1},\ldots,k_{i_l}=-\infty}^{\infty} \alpha_D(S;k_{i_1},\ldots,k_{i_l}) \left(1-\cos (k_{i_1}\beta_{i_1}+\ldots+k_{i_l}\beta_{i_l}) \right) \right]} \nonumber \\
                                                              &     & \prod_{j=1}^n \exp{\left[-\lambda \sum_{k=-\infty}^{\infty} \alpha_{\delta,D}(z_j;k)(1 - \cos(k\beta_j)) \right]} \, ,
\end{eqnarray*}
For each $S \subset \{ z_1, \ldots, z_n \}$ with $|S| = l \geq 2$, $\alpha_D(S;k_{i_1},\ldots,k_{i_l})$ is invariant under conformal transformations; therefore,
\begin{equation} \nonumber
\frac{\left\langle e^{i \beta_1 N_{w}(z_1)} \ldots e^{i \beta_n N_{w}(z_n)} \right\rangle_{\delta,D}}{\left\langle e^{i \beta_1 N_{w}(z'_1)} \ldots e^{i \beta_n N_{w}(z'_n)} \right\rangle_{\delta,D'}}
= \prod_{j=1}^n \exp{\left\{-\lambda \sum_{k=-\infty}^{\infty} \left[\alpha_{\delta,D}(z_j;k)-\alpha_{\delta,D'}(z'_j;k)\right](1 - \cos(k\beta_j)) \right\}} \, . 
\end{equation}
Proceeding as in the proof of Theorem \ref{thm-bounded-domains}, but using Lemma~\ref{SecondLemma}
instead of Lemma~\ref{FirstLemma}, gives
\begin{equation} \nonumber
\alpha_{\delta,D}(z_j;k)-\alpha_{\delta,D'}(z'_j;k) = -c_k \log|f'(z_j)| + o(1) \text{ as } \delta \to 0 \, ,
\end{equation}
where $c_k = \frac{1}{2\pi^2 k^2}$ for $k \in {\mathbb Z} \setminus \{0\}$ and $c_0 = 1/30$.

This, together with the observation, already used at the end of Section~\ref{Sec:WindingOperator}, that
$\sum_{k=-\infty}^{\infty} c_k (1-\cos(k\beta)) = \frac{\beta(2\pi-\beta)}{4\pi^2}$ (where, on the right hand side,
$\beta$ should be interpreted modulo $2\pi$), readily implies the statement of the theorem. \fbox{}

\subsection{The layering model in the plane} \label{n-point-plane}

Recall that $\left\langle \prod_{j=1}^n V_{\beta_j}(z_j) \right\rangle_{\delta,R}$ denotes
the expectation of the product $\prod_{j=1}^n e^{i \beta_j N_\ell(z_j)}$ with respect to a loop soup in $\mathbb C$
with intensity $\lambda>0$ containing only loops $\gamma$ of diameter $0<\delta\leq\diam(\gamma)<R<\infty$.
\begin{theorem} \label{thm-n-point-plane}
If $n \in {\mathbb N}$ and $\boldsymbol{\beta}=(\beta_1,\ldots,\beta_n)$
with $|\boldsymbol{\beta}| = \sum_{j=1}^n \beta_j \in 2\pi {\mathbb Z}$, then 
\begin{equation} \nonumber
\lim_{\delta \to 0, R \to \infty} 
\frac{\left\langle \prod_{j=1}^n V_{\beta_j}(z_j) \right\rangle_{\delta, R}}{\delta^{ \frac{\lambda}{5}\sum_{j=1}^n (1-\cos\beta_j)}} \equiv \phi_{\mathbb C}(z_1, \dots, z_n;\boldsymbol{\beta}) 
\end{equation}
exists and is finite and real. Moreover, if $f:{\mathbb C} \to {\mathbb C}$ is a conformal map such that
$z'_1=f(z_1),\ldots,z'_n=f(z_n)$, then
\begin{equation} \nonumber
\phi_{\mathbb C}(z'_1,\ldots,z'_n;\boldsymbol{\beta}) = \prod_{j=1}^n \left|f'(z_j)\right|^{-\frac{\lambda}{5}(1-\cos\beta_j)} \phi_{\mathbb C}(z_1, \dots, z_n;\boldsymbol{\beta}) \, .
\end{equation}
\end{theorem}

\noindent{\bf Proof sketch.} The beginning of the proof proceeds like that of Theorem~\ref{thm-bounded-domains} until equation~\eqref{n-point-func},
leading to the following equation:
\begin{eqnarray*}
\left\langle \prod_{j=1}^n V_{\beta_j}(z_j) \right\rangle_{\delta,R} & = &
\prod_{S \subset \{z_1,\ldots,z_n \}, 1<|S|<n} \exp{\left[-\lambda \alpha_R(S) \left(1-\cos\left(\sum_{k \in I_S} \beta_k \right)\right)\right]} \\
& & \prod_{j=1}^n \exp{\left[-\lambda \alpha_{\delta,R}(z_j)(1 - \cos \beta_j) \right]} \, ,
\end{eqnarray*}
where $\alpha_R(S) \equiv \mu_{\mathbb C}(\gamma: S \subset \bar \gamma, S^c \subset \bar\gamma^c, \diam(\gamma)<R)$,
for $S \subset \{ z_1, \ldots, z_n \}$ with $2\leq|S|<n$, and
$\alpha_{\delta,R}(z_j) \equiv \mu_{\mathbb C}(\gamma: \delta \leq \diam(\gamma) < R, z_j \in \bar \gamma, z_k \notin \bar\gamma \; \forall k \neq j)$,
and where $I_S$ denotes the set of indices such that $k \in I_S$ if and only if $z_k \in S$.

Note, in the equation above, the condition $|S|<n$ in the first product on the right hand side; this condition comes from the fact that
the term $-\lambda \alpha_R(S)$ with $S = \{z_1,\ldots,z_n\}$ is multiplied by $1-\cos(\sum_{k=1}^n \beta_k)=0$, where we have
used the ``charge conservation'' condition $|\boldsymbol{\beta}| = \sum_{j=1}^n \beta_j \in 2\pi {\mathbb Z}$.

For every $j=1,\ldots,n$, using Lemma~\ref{FirstLemma}, we have that
 \begin{eqnarray*}
  \alpha_{\delta,R}(z_j) &=& \mu_{\mathbb C}(\gamma: \delta \leq \diam(\gamma) < R, z_j \in \bar \gamma, z_k \notin \bar\gamma \; \forall k \neq j)  \\
  & = & \mu_{\mathbb C}(\gamma: m > \diam(\gamma) \geq \delta, z_j \in \bar \gamma) \\
  & + & \mu_{D}(\gamma: m\leq\diam(\gamma)<R, z_j \in \bar \gamma, z_k \notin \bar\gamma \; \forall k \neq j)\\
  &=&
  \frac{1}{5} \log\frac{m}{\delta} + \alpha_{m,R}(z_j).
 \end{eqnarray*}
Now note that monotonicity and the fact that the Brownian loop soup is \emph{thin} \cite{2010arXiv1009.4782N} imply that
$\alpha_{m,{\mathbb C}}(z_j) \equiv \lim_{R \to \infty} \alpha_{m,R}(z_j)$ and $\alpha_{\mathbb C}(S) \equiv \lim_{R \to \infty} \alpha_R(S)$,
for $S \subset \{ z_1, \ldots, z_n \}$ with $2\leq|S|<n$, exist and are bounded.
After letting $R\to\infty$, the proof proceeds like that of Theorem~\ref{thm-bounded-domains}, with $D=D'=\mathbb C$. \noindent \fbox{} \\

We have already seen the behavior of the 2-point function in the layering model in Section~\ref{2-point-function-layering-model}; the theorem below
deals with the 3-point function.
\begin{theorem} \label{3-point-function}
Let $z_1, z_2, z_3 \in {\mathbb C}$ be three distinct points, then we have that
\begin{eqnarray*}
\lefteqn{
\phi_{\mathbb C}(z_1,z_2,z_3;\beta_1,\beta_2,\beta_3) = } \\
& & C_{3} \left| \left( {1 \over |z_1 - z_2|} \right)^{\Delta_{l}(\beta_1) + \Delta_{l}(\beta_2) - \Delta_{l}(\beta_3)}  \left( {1 \over |z_1 - z_3|} \right)^{\Delta_{l}(\beta_1) + \Delta_{l}(\beta_3) - \Delta_{l}(\beta_2)} 
 \left( {1 \over |z_2 - z_3|} \right)^{\Delta_{l}(\beta_2) + \Delta_{l}(\beta_3) - \Delta_{l}(\beta_1)} \right|^2
\end{eqnarray*}
for some constant $C_{3}$.
\end{theorem}

\noindent{\bf Proof.}
Theorem~\ref{thm-n-point-plane} implies that the 3-point function in the full plane transforms covariantly under
conformal maps. This immediately implies the theorem following standard argument (see, {\it e.g.}, \cite{MR1424041}).
We briefly sketch those arguments below for the reader's convenience.

Scale invariance, rotation invariance, and translation invariance immediately imply that there are constants $C_{abc}$ such that
\be \label{eq-invariance}
\phi_{\mathbb C}(z_1,z_2,z_3;\beta_1,\beta_2,\beta_3) = \sum C_{abc} z_{12}^{-a} z_{13}^{-b} z_{23}^{-c} \, ,
\ee
where $z_{ij}=|z_i-z_j|$ and the sum is over all triplets $a,b,c \geq 0$ satisfying $a+b+c = 2(\Delta_l(\beta_1) + \Delta_l(\beta_2) + \Delta_l(\beta_3))$
(the constraint on the exponents $a,b,c$ follows from Theorem~\ref{thm-n-point-plane} applied to scale transformations).

Now let $f$ be a conformal transformation from $\mathbb C$ to $\mathbb C$; $f$ is then a M\"obius transformation and has the form
$f(z) = \frac{Az+B}{Cz+D}$, with $f'(z) = \frac{AD-BC}{(Cz+D)^2}$. Letting $\gamma_j \equiv |f'(z_j)|^{-1}$, if $\tilde z = f(z)$, it is easy
to check that $\tilde z_{ij} = \gamma_i^{-1/2} \gamma_j^{-1/2} z_{ij}$. Using this fact and Theorem~\ref{thm-n-point-plane}, we have that
\begin{eqnarray*}
\phi_{\mathbb C}(\tilde z_1, \tilde z_2, \tilde z_3;\beta_1,\beta_2,\beta_3) & = &
\left(\gamma_1^{\Delta_l(\beta_1)} \gamma_2^{\Delta_l(\beta_2)} \gamma_3^{\Delta_l(\beta_3)}\right)^2 \sum C_{abc} z_{12}^{-a} z_{13}^{-b} z_{23}^{-c} \\
& = & \left(\gamma_1^{\Delta_l(\beta_1)} \gamma_2^{\Delta_l(\beta_2)} \gamma_3^{\Delta_l(\beta_3)}\right)^2
\sum C_{abc} \frac{\tilde z_{12}^{-a} \tilde z_{13}^{-b} \tilde z_{23}^{-c}}{\gamma_1^{a/2+b/2} \gamma_2^{a/2+c/2} \gamma_3^{b/2+c/2}} \, .
\end{eqnarray*}
For this last expression to be of the correct form~\eqref{eq-invariance}, the $\gamma$'s need to cancel; this immediately leads to the relations
$a = 2(\Delta_l(\beta_2) + \Delta_l(\beta_3) - \Delta_l(\beta_1))$, $b = 2(\Delta_l(\beta_1) + \Delta_l(\beta_3) - \Delta_l(\beta_2))$,
$c = 2(\Delta_l(\beta_1) + \Delta_l(\beta_2) - \Delta_l(\beta_3))$. \fbox{}

\section{Dimension of the exponential winding operator}\label{dimexp}

In this section we compute the conformal dimension of the exponential winding number operator
$e^{i \beta N_{w}(z)}$ for the BLS in the plane using non-rigorous path integral methods.

The measure for a single Brownian loop of time length $t$ rooted at $x$ is given by the path integral for  a free particle in two Euclidean dimensions, with the path $y(\tau)$ constrained to begin and end at $x$:
\be \label{spart}
 \int_{y(0)=x}^{y(t)=x} [d^{2}y] e^{-{1 \over 2}\int_{0}^{t} d\tau~ |\dot y|^{2}},
\ee
 where $\dot y \equiv \p y/\p \tau$, $y,x$ are complex coordinates, and the path integral measure is Brownian normalized so that $ \int_{x}^{z} [d^{2}y] e^{-{1 \over 2}\int_{0}^{t} d\tau~ |\dot y|^{2}}=e^{-|x-z|^{2}/2t}/(2 \pi t)$.

To construct the Brownian loop soup in the plane, we should integrate the root point $x$ over the plane with uniform measure, integrate over $t$ with measure $dt/t$, and sum over $n$-loop sectors  weighted by $\lambda^{n}$ and divided by $n!$ since the loops are indistinguishable:
\be \label{part}
\begin{split}
Z_{\lambda} = 1 + \sum_{n=1}^\infty {\lambda^n \over n!} \prod_{k=1}^n \int_{\delta^{2}}^{R^{2}} {d t_k \over t_k} \int d^2x_k  \int_{y_{k}(0)=x_{k}} ^{y_{k}(t_{k})=x_{k}} [d^{2}y_{k}] \exp\({-{1 \over 2}\int_{0}^{t_{k}} d\tau_{k}~ |\dot y_{k}|^{2}}\) \\ = \exp
  \( \lambda \int_{\delta^2}^{R^2} {d t \over t} \int d^2x   \int_{y(0)=x}^{y(t)=x} [d^{2}y] e^{-{1 \over 2}\int_{0}^{t} d\tau~ |\dot y|^{2}} \) .
\end{split}
\ee
This result is closely related to (3.1) of \cite{Freivogel:2009rf}.

We wish to compute the 1-point function  $\la e^{i \beta N_{w}(z)} \ra,$
where $\la ... \ra$ represents the average over the distribution defined by \eqref{part}.   Since $N_{w}$ is  integer the result should be invariant under $\beta \rightarrow \beta+2 \pi n$.  Furthermore for every configuration there is a mirror configuration where $N \to -N$, so $\la e^{i \beta N(z)} \ra$ is real and invariant under $\beta \rightarrow -\beta$.  Because we are in the plane and charge conservation cannot be satisfied, the 1-point function will diverge as a power of the ratio of the cutoffs $R/\delta$, but the power  tells us the dimension of the operator we wish to compute (\emph{cf.}  \eqref{winddim}).

To compute  $\la e^{i \beta N_{w}(z)} \ra$ we  can insert $ e^{i \beta N_w(z)}$ in the path integral for each loop in \eqref{part}.  By translation invariance in the plane  we can set $z=0$.  Then this insertion corresponds to adding a term $i \beta \int_{0}^{t} d\tau~\dot \phi(\tau)/2 \pi$ to the single-particle ``lagrangian'' in the exponential of \eqref{spart}, where $\phi(\tau) \,\, {\rm mod} \,\, 2 \pi = \arg(y(\tau))$ is the angular position of the path $y(\tau)$ relative to the origin $z=0$.
Therefore
\be \nonumber
\la e^{i \beta N_{w}(z)} \ra = Z_{\lambda}^{-1}~ \exp
  \( \lambda  \int_{\delta^{2}}^{R^{2}} {d t \over t} \int d^2x   \int_{y_(0)=x}^{y(t)=x} [d^2 y] e^{\int_{0}^{t} d\tau \[ -{1 \over 2} (\dot y)^{2} + i \beta \dot \phi/2 \pi \]} \) .
\ee
Apart from the integral over $t$, the quantity in the exponential is related to the canonical partition function at inverse temperature $t$ of a charged particle in the field of a single magnetic monopole of charge $\beta/2 \pi$ at the origin ($n$-point functions involving products of exponentials $e^{i \beta_{n} N_{w}(z_{n})}$ would  correspond to multiple magnetic monopoles with charge $\beta_{n}/2 \pi$ at the locations $z_{n}$).  To compute it, one can  quantize the Hamiltonian for a particle in the monopole field, then take the trace in the energy basis \cite{Arovas:1985yb}, or perform the position-space path integral directly and obtain a result in terms of a sum over Bessel functions \cite{PhysRevD.20.2550}.  The final result is (see for instance \cite{2011JSMTE..05..024D}):
$$
 \int d^2x   \int_{y(0)=x}^{y(t)=x} [d^2 y] e^{-\int_{0}^{t} d\tau  {1 \over 2} (\dot y)^{2}} \(e^{i \beta \int_{0}^{t} d\tau  \dot \phi /2 \pi}- 1 \) = - \beta (2 \pi-\beta)/8 \pi^{2}
$$
valid for $0\leq \beta < 2 \pi$ and periodic in $\beta \simeq \beta + 2 \pi n$.  From this we obtain
\be \label{physwind}
\left\la e^{i \beta N_{w}(z)} \right\ra = \exp \( - \lambda  {  \beta (2 \pi-\beta) \over 8 \pi^2} \int_{\delta^{2}}^{R^{2}} {d t \over t}  \) = \({R \over \delta } \)^{ -  \lambda   \beta (2 \pi-\beta)/4 \pi^{2}}
\ee
 in agreement with \eqref{winddim}.

\section{Central charge and relation to the free boson} \label{sec-central-charge}

The partition function \eqref{part} has a simple interpretation for the  case $\lambda = 1/2$.  It corresponds precisely to the partition function for a free, massless, real bosonic field in two (Euclidean) dimensions:
$$\ln Z_{\rm  boson} = - {1 \over 2} \log \det H = {1 \over 2} \int_0^\infty {dt \over t} \, {\rm Tr} e^{-t H} =  {1 \over 2} \int_0^\infty {dt \over t} \int d^2 x \, \left\langle x | e^{-t H} | x \right\rangle,$$
where $H = \Box$ is the Laplace operator.  Expressing the heat kernel $\left\langle x | e^{-t H} | y \right\rangle$ using the path integral \eqref{spart} completes the identification with \eqref{part}.

An analogous relation between the partition function of the discrete Gaussian free field (the lattice analog of the free field)
and that of the random walk loop soup (the lattice analog of the BLS) can be derived easily; namely $Z^{D}_{DGFF} = \left(\frac{\pi}{2}\right)^{|D|} Z^{D}_{1/2}$, where
$Z^{D}_{DGFF}$ is the partition function of the discrete Gaussian free field in $D$ with zero boundary condition, $Z^{D}_{1/2}$ is the partition function of the random
walk loop soup in $D$ with intensity $1/2$, and $|D|$ is the number of vertices in $D$ (see, for example \cite{2015arXiv150104861C}, particularly Sections 2.1 and 2.2
and Exercise 2.5). This result may be related to work of Le Jan, who demonstrated that for $\lambda = k/2$, the occupation field for the BLS can be identified with the
sum of squares of $k$ copies of a free field \cite{MR2815763}. It suggests that for general $\lambda$, the BLS might give meaning to the notion of a fractional power of
a free field.  

Because the massless boson is a CFT with central charge $c=1$, the central charge of the BLS appears to be $c=2 \lambda$ for the case $\lambda = 1/2$.
Given the form of \eqref{part}, one can express the partition function of the BLS for arbitrary $\lambda$ in terms of that for $\lambda=1/2$:
$Z_{\lambda} = \left( Z_{1/2} \right)^{2\lambda}$. The same relation can be rigorously proved to hold for the random walk loop soup mentioned earlier
(see equation~(2.2) in Section~2.1 of~\cite{2015arXiv150104861C}).
This leads to the conclusion that the relation $c(\lambda)=2\lambda$ between the intensity $\lambda$ and the central charge $c$ of the BLS holds for all $\lambda$.

This agrees with the central charge of the SLE corresponding to the ensemble of BLS cluster boundaries.
Indeed, conformal field theory considerations lead to
the formula
\be \nonumber
c(\kappa) = \frac{(3\kappa-8)(6-\kappa)}{2\kappa} ,
\ee
but it is also known~\cite{MR2979861} that, when $8/3 < \kappa \leq 4$, the ensemble of cluster boundaries of a BLS with intensity
\be \nonumber
\lambda = \frac{(3\kappa-8)(6-\kappa)}{4\kappa} = \frac{c(\kappa)}{2}
\ee
is a CLE$_\kappa$. (For example, a BLS with $\lambda = 1/4$ gives a CLE$_3$, with central charge $1/2$.)

We note that most of the existing literature, including~\cite{MR2979861}, contains an error in the correspondence between $\kappa$
and the loop soup intensity $\lambda$. The error can be traced back to the choice of normalization of the (infinite) Brownian loop
measure $\mu^{loop}$, which determines the constant in front of the log in Lemma~\ref{FirstLemma} below.\footnote{We thank Greg
Lawler for discussions on this topic.} With the normalization used in this paper, which coincides with the one in the original definition of the
Brownian loop soup \cite{2003math4419L}, for a given $8/3 < \kappa \leq 4$, the corresponding value of the loop soup intensity $\lambda$
is half of that given in~\cite{MR2979861}.

\section{Conclusions} \label{conclusions}

There is still only a partial understanding of the rich
connection between the conformal field theories studied by physicists and conformally invariant stochastic models such as the BLS or SLE.
Physicists are  often interested in CFTs defined by a Lagrangian, and the correlation functions of primary operators, the spectrum of conformal dimensions, and the central charge are the main objects of study and interest.  By contrast, conformal stochastic models are often defined and studied using very different methods and with different goals.

Here we have tried to take a few steps towards strengthening this connection. Using rigorous methods, we defined a set of quantities in the BLS and proved that their expectation values behave like the correlation functions of primary operators.  While we have not established it, we expect that these correlation functions may define a conformal field theory.  If so, it has several novel features, such as a periodic spectrum of conformal dimensions.

There are many basic questions that remain to be answered.  Does this approach to the BLS in fact define a CFT?  If so, have we found the complete set of primary operators?  What is the stress-energy tensor? Is the theory reflection positive and/or modular invariant?  Is it unique in some sense?

An even more ambitious set of questions relates to eternal inflation, albeit in 2+1 dimensions.  The model of \cite{Freivogel:2009rf} was proposed as  a toy model for the late-time evolution of an eternally inflating spacetime, but the lack of analyticity in the 4-point function of the (analog of the) layering operator derailed it as a putative CFT.  Does the BLS solve this problem, and if so, could it help define a CFT dual to eternal inflation?  What type of object in 2+1 dimensional de Sitter space produces the BLS as its late-time distribution?  Is there a natural generalization to higher dimensions?

\section{\bf Appendix. The Brownian loop measure: two lemmas} \label{appendix}

In this appendix we prove two important lemmas which are used several times in the rest of the paper. The lemmas concern
the $\mu^{loop}$-measures of certain sets of loops, where, as in the rest of the paper, $\mu^{loop}$ is the intensity measure
used in the definition of the Brownian loop soup (see equation~\eqref{brownian-loop-measure}).

\begin{lem} \label{FirstLemma}
Let $z \in \mathbb C$, then
\be \nonumber
\mu^{loop}(\gamma: z \in \bar \gamma, \delta \leq \diam(\gamma) < R)
= \frac{1}{5} \log\frac{R}{\delta} .
\ee
\end{lem}

\noindent{\bf Proof.}
Since $\mu^{loop}(\gamma: \diam(\gamma)=R)=0$, we have that
\begin{eqnarray*}
\lefteqn{\mu^{loop}(\gamma: z \in \bar \gamma, \delta \leq \diam(\gamma) < R)
- \mu^{loop}(\gamma: z \in \bar \gamma,\gamma \not\subset B_{z, \delta}, \gamma \subset B_{z,R})} \\
& = & \mu^{loop}(\gamma: z \in \bar \gamma, \diam(\gamma) \geq \delta, \gamma \subset B_{z,\delta})
- \mu^{loop}(\gamma: z \in \bar \gamma,\diam(\gamma) \geq R, \gamma \subset B_{z,R}) ,
\end{eqnarray*}
where $B_{z, a}$ is a disk of radius $a$ around $z$, and $\gamma \not\subset B_{z, \delta}$ indicates that the image of $\gamma$ is not fully contained in $B_{z, \delta}$.
By the scale invariance of $\mu^{loop}$, the last two terms are identical, so that
\begin{eqnarray} \label{log-equation}
\mu^{loop}(\gamma: z \in \bar \gamma, \delta \leq \diam(\gamma) < R) & = &
\mu^{loop}(\gamma: z \in \bar \gamma,\gamma \not\subset B_{z, \delta}, \gamma \subset B_{z,R}) \nonumber \\
& = & c \log\frac{R}{\delta}
\end{eqnarray}
for some positive constant $c<\infty$, where the last equality follows from Proposition 3 of \cite{2005math.....11605W} and the fact that the Brownian loop measure satisfies the conformal restriction property.
In order to determine the constant $c$, we use the fact that, for any $r>0$,
\begin{equation} \label{measure-equality}
\mu^{loop}\left(\gamma: z \in \bar \gamma, 1 \leq \diam(\gamma) < e^r \right)
= \mu^{loop}\left(\gamma: z \in \bar \gamma, 1 \leq t_{\gamma} < e^{2r}\right) .
\end{equation}
(Equation \eqref{measure-equality} is essentially a consequence of Brownian scaling and can be proved using standard techniques.
The interested reader can consult, for example, Appendix B of \cite{2015arXiv150104861C}).

Following \cite{2005math.....11605W}, we compute the right hand side of \eqref{measure-equality} using the definition of Brownian loop measure
and translation invariance:
\begin{eqnarray*}
\mu^{loop}(\gamma: 0 \in \bar \gamma, 1 \leq t_{\gamma} < e^{2r})
& = & \int_{\mathbb C} \int_1^{e^{2r}} \frac{1}{2 \pi t^2} \,
\mu^{br}_{z,t}(\{ \gamma: 0 \in \bar \gamma \}) \, dt \, d{\bf A}(z) \\
& = & \int_{\mathbb C} \int_1^{e^{2r}} \frac{1}{2 \pi t^2} \,
\mu^{br}_{0,t}(\{ \gamma: z \in \bar \gamma \}) \, dt \, d{\bf A}(z) \\
& = & \int_1^{e^{2r}} \frac{1}{2 \pi t^2} \,
{\mathbb E}^{br}_{0,t}\left( \int_{\mathbb C} \mathbbm{1}_{\{ \gamma: z \in \bar \gamma \}} \, d{\bf A}(z) \right) \, dt \\
& = & \int_1^{e^{2r}} \frac{1}{2 \pi t} \,
{\mathbb E}^{br}_{0,1}\left( \int_{\mathbb C} \mathbbm{1}_{\{ \gamma: z \in \bar \gamma \}} \, d{\bf A}(z) \right) \, dt ,
\end{eqnarray*}
where ${\mathbb E}^{br}_{0,t}$ denotes expectation with respect to a complex Brownian bridge of time length $t$
started at the origin, and where, in the last equality, we have used the fact that
\begin{equation} \nonumber
{\mathbb E}^{br}_{0,t}\left( \int_{\mathbb C} \mathbbm{1}_{\{ \gamma: z \in \bar \gamma \}} \, d{\bf A}(z) \right)
= t \, {\mathbb E}^{br}_{0,1}\left( \int_{\mathbb C} \mathbbm{1}_{\{ \gamma: z \in \bar \gamma \}} \, d{\bf A}(z) \right)
\end{equation}
because of scaling. The expected area of a ``filled-in'' Brownian bridge, computed in \cite{2006CMaPh.264..797G}, is
\begin{equation} \nonumber
{\mathbb E}^{br}_{0,1}\left( \int_{\mathbb C} \mathbbm{1}_{\{ \gamma: z \in \bar \gamma \}} \, d{\bf A}(z) \right)
= \frac{\pi}{5} ,
\end{equation}
so that
\begin{equation} \label{frac-equation}
\mu^{loop}(\gamma: z \in \bar \gamma, 1 \leq t_{\gamma} < e^{2r}) = \frac{r}{5} .
\end{equation}
Using \eqref{frac-equation} and \eqref{measure-equality}, we obtain
\begin{equation} \nonumber
\mu^{loop}(\gamma: z \in \bar \gamma, 1 \leq \diam(\gamma) < e^r) = \frac{r}{5} .
\end{equation}
Comparing this to \eqref{log-equation} gives $c=1/5$. \fbox{} \\

\begin{lem} \label{SecondLemma}
Let $z \in \mathbb C$ and $k \in {\mathbb Z} \setminus \{0\}$, then
\be \nonumber
\mu^{loop}(\gamma: \gamma \text{ has winding number } k \text{ around } z, \delta \leq \diam(\gamma) < R)
= \frac{1}{2\pi^2 k^2} \log\frac{R}{\delta} .
\ee
\end{lem}

\noindent{\bf Proof.} It is easy to check that the measure on loops surrounding the origin induced by $\mu^{loop}$,
but restricted to loops that wind $k$ times around a given $z \in {\mathbb C}$, satisfies the conformal restriction
property. Therefore, as in the proof of Lemma~\ref{FirstLemma}, we have
\begin{eqnarray*}
\lefteqn{\mu^{loop}(\gamma: \gamma \text{ has winding number } k \text{ around } z, \delta \leq \diam(\gamma) < R)} \\
& = & \mu^{loop}(\gamma: \gamma \text{ has winding number } k \text{ around } z, \gamma \not\subset B_{z, \delta}, \gamma \subset B_{z,R}) \nonumber \\
& = & c_k \log\frac{R}{\delta} ,
\end{eqnarray*}
for some positive constant $c_k<\infty$, where in the last equality we have used Proposition 3
of~\cite{2005math.....11605W}.
In order to find the constants $c_k$, we can proceed as in the proof of Lemma~\ref{FirstLemma}, using
the fact that the expected area of a ``filled-in'' Brownian loop winding $k$ times around the origin was
computed in \cite{2006CMaPh.264..797G} and is equal to $1/2\pi k^2$ for $k \in {\mathbb Z} \setminus \{0\}$
(and $\pi/30$ for $k=0$). \fbox{}

\section*{Acknowledgements}
It is a pleasure to thank M.~Bauer, D.~Bernard, B.~Duplantier, B.~Freivogel, P.~Kleban, G.~Lawler, T.~Lupu, Y.~Le Jan, M.~Lis, M.~Porrati, A.~Sokal, and S.~Storace for discussions.
The work of FC is supported in part by the Netherlands Organization for Scientific Research (NWO) through grant Vidi 639.032.916.
The work of MK is supported in part by the NSF  through grant PHY-1214302.

\bibliography{blsbib2}

\vskip .5cm
\vskip .5 cm

\end{document}